\documentclass[12pt,preprint]{aastex}
\usepackage{epsfig,natbib}
\usepackage{emulateapj5}
\input psfig.tex

\shortauthors{Eracleous, Halpern, \& Charlton}
\shorttitle{X-Ray Spectrum and UV Absober of Arp 102B}

\newcounter{species}
\def\ion#1#2{\setcounter{species}{#2}#1$\;${\sc\roman{species}}\relax}
\def\lion#1#2#3{\setcounter{species}{#2}#1$\;${\sc\roman{species}$\;\lambda${#3}}\relax}
\def\llion#1#2#3{\setcounter{species}{#2}#1$\;${\sc\roman{species}$\;\lambda\lambda${#3}}\relax}

\def\kms{\ifmmode{~{\rm km~s^{-1}}}\else{~km s$^{-1}$}\fi}
\def\lsim{\lower0.3em\hbox{$\,\buildrel <\over\sim\,$}}
\def\gsim{\lower0.3em\hbox{$\,\buildrel >\over\sim\,$}}

\def\asca{{\it ASCA}}
\def\chandra{{\it Chandra}}
\def\xmm{{\it XMM-Newton}}
\def\lam{$\lambda$}

\begin{document}

\title{The ASCA X-Ray Spectrum of Arp 102B and Evaluation of Simple
Models for Its Associated, Metastable Fe~II Absorber}

\author{Michael Eracleous\altaffilmark{1}, 
Jules P. Halpern\altaffilmark{2}, 
\& Jane C. Charlton\altaffilmark{1}}

\altaffiltext{1}{Department of Astronomy and Astrophysics, The
Pennsylvania State University, 525 Davey Lab, University Park, PA
16802}

\altaffiltext{2}{Department of Astronomy Columbia University, 538 West
120th Street, New York, NY 10027}

\begin{abstract}

We have observed the broad-line radio galaxy Arp~102B with {\it ASCA}
in order to determine the absorbing column density towards its X-ray
source and measure its X-ray spectrum. The ultimate goal was to
constrain the properties of the medium responsible for the metastable
\ion{Fe}{2} absorption lines observed in the ultraviolet spectrum of
this object. The 0.5--10~keV X-ray spectrum is best described by a
simple power-law model of photon index $1.58\pm0.04$ modified by
photoelectric absorption with an equivalent hydrogen column density of
$(2.8\pm0.3)\times 10^{21}~{\rm cm}^{-2}$. An Fe~K$\alpha$ line is not
detected with an upper limit to its equivalent width of 200~eV,
assuming that its full width at half maximum is 60,000\kms.  Using the
column density measured from the X-ray spectrum and the observed
spectral energy distribution as constraints, we explore simple
(single-zone, constant-density) photoionization models for the
absorber for a wide range of densities and ionization parameters in an
effort to reproduce the strengths of the ultraviolet absorption
lines. We find that densities of at least $10^{11}~{\rm cm^{-3}}$ are
needed. However, a single ionization parameter cannot explain all of
the observed lines. An ionization parameter between $10^{-2.5}$ and
$10^{-3.5}$ is needed to explain the Mg and Fe lines and the soft
X-ray absorption, but the observed lines (from Si, C, Al, and H)
require different density--ionization parameter
combinations. According to the models, such an absorbing medium must
be located very close to the source of ionizing radiation (within
5,000 gravitational radii) and must be very compact. As such, the
properties of this absorbing medium differ from those of more luminous
quasars, but are reminiscent of the absorber in the Seyfert galaxy
NGC~4151. We suggest that the absorber is in the form of thin sheets
or filaments embedded in an outflowing wind that overlays the
accretion disk of Arp~102B. This picture is consistent with all of the
available constraints on the central engine of this object. In an
appendix, we present the X-ray spectrum of the source MS~1718.6+4902,
which happened to fall within the field of view of the {\it ROSAT}
PSPC and the {\it ASCA} GIS during the observations of Arp~102B.

\end{abstract}

\keywords{galaxies: active---galaxies:individual (Arp 102B,
MS~1718.6+4902)--- X-rays: galaxies---line: formation}


\section{Introduction}

The broad-line radio galaxy (BLRG) Arp~102B \citep*{ssk83,chf89,ch89}
is the prototype of a subset of radio-loud active galactic nuclei
(AGNs) with double-peaked Balmer lines in their optical spectra
\citep[see more examples in][]{eh94}. The profiles of these
double-peaked emission lines and a number of additional, extreme
spectroscopic properties of these objects have led to the
interpretation that they harbor line emitting accretion disks whose
inner parts have the form of an optically thin, geometrically thick ion
torus \citep{ch89,eh94}. The ion torus \citep{r82} is very similar to
what is known today as an advection-dominated accretion flow (ADAF;
Narayan \& Yi 1994, 1995\nocite{ny94,ny95}), a structure which is
realizable at accretion rates considerably lower than the Eddington
rate. Although alternative scenarios for the origin of the
double-peaked lines have been proposed, these are disfavored by the
observational evidence available today (see reviews by Eracleous 1998,
1999\nocite{e98,e99}).

In the process of studying the UV spectroscopic properties of AGNs
with double-peaked emission lines we obtained a spectrum of Arp~102B
with the {\it Hubble Space Telescope (HST)}, which revealed an
unexpected complex of \ion{Fe}{2} {\it absorption} lines very close to
its systemic redshift \citep{h96}. This complex includes much more
than the transitions from the ground state of \ion{Fe}{2} that are
often observed in quasar absorption line systems and the interstellar
medium (multiplets UV1, UV2, and UV3 from the $a\,^6D$ ground state
term): it also includes multiplets from the next two lowest states
(terms $a\,^4F$ and $a\,^4D$).  Particularly surprising is the
presence of multiplets UV62, UV63, and UV64, arising from the
metastable $a\,^4D$ term, which lies 0.99--1.10~eV above the ground
state. Since absorption from excited states of \ion{Fe}{2} is
incompatible with the diffuse interstellar medium of the host galaxy
and with intervening absorption-line systems, the absorber is most
likely intrinsic to Arp~102B and its association with the active
nucleus remains to be understood. Metastable \ion{Fe}{2} absorption
lines are not observed in Seyfert galaxies very often, with NGC~4151
being one of the rare examples \citep{kraemer01}. NGC~4151 also shows
a \ion{C}{3}]~\lam1176 line, arising from a level 6.05~eV above the
ground state \citep{k95}. Absorption lines from the ground state of
\ion{Fe}{2} have been observed in some other Seyfert galaxies, for
example, NGC~3227 \citep{crenshaw01}.  In marked contrast with Seyfert
galaxies, \ion{Fe}{2} absorption lines from metastable states are
often seen in the spectra of quasars. Examples that have been known
for quite some time include Q~0059--2735 \citep*{h87,wcp95},
Hawaii~167 \citep{c94}, and Mrk~231 \citep*{bmm91,s95}. Recently, a
number radio-loud quasars from the FIRST radio survey \citep*{bwh95}
have been found to host metastable \ion{Fe}{2} lines as well
\citep{b97,b00}, as well as a number of quasars from the Sloan Survey
\citep{hall02}. Finally, we ourselves have found systems of metastable
\ion{Fe}{2} absorption lines, identical to that in Arp~102B, in the
BLRG 3C~332 and in the LINER NGC~1097 \citep{h97,e02}.

The excitation state of the gas responsible for the \ion{Fe}{2}
absorption lines in Arp~102B is very similar to what is observed in
Q~0059--2735. The absorber in the latter object has been studied in
detail by \citet{wcp95}, who suggested that it is made up of
condensations in a hotter broad-absorption line flow. The temperature
of the condensations is likely to be of order $10^4$~K, implying that,
if the metastable \ion{Fe}{2} levels are populated by collisions, the
electron density should be greater than $10^6~{\rm cm}^{-3}$.  In our
effort to understand the nature of the absorber in Arp~102B, we
studied the X-ray spectrum obtained with the {\it ROSAT} PSPC and
found a very large absorbing column of neutral gas, in the range
$(2-8)\times 10^{21}~{\rm cm}^{-2}$ \citep{h97}. Because the {\it
ROSAT} spectrum was so heavily absorbed, more stringent constraints
could not be derived. Therefore, we observed Arp~102B with {\it
ASCA}. This paper is devoted to the analysis and interpretation of the
{\it ASCA} spectra, including the development and assessment of simple
photoionization models for the absorbing medium. Preliminary results
of this work were presented by \citet{e02}.  In \S2 we describe the
observations and data screening and in \S3 we present the light curve
and discuss the time variability. In \S4 we study the X-ray continuum
and determine the column density of the absorbing medium, while in \S5
we derive upper limits to the equivalent width ($EW$) of the
Fe~K$\alpha$ line. In \S6 we construct and evaluate simple
photoionization models for the medium responsible for the UV and X-ray
absorption.  We discuss our findings in \S7 where we examine the X-ray
spectral properties of Arp~102B in the context of other BLRGs and
consider the implications of the model results for the nature of the
absorber.  In an appendix we present the X-ray spectrum of
MS~1718.6+4902, which happens to be in the field of view of
Arp~102B. Throughout this paper we assume a Hubble constant of
50~km~s$^{-1}$~Mpc$^{-1}$, which implies a distance to Arp~102B of
146~Mpc, given its redshift of $z=0.0244$.

\section{ASCA Observations and Data Screening}

Arp~102B was observed with {\it ASCA} on 1998 February 19 with the SIS
detectors in {\tt 1-CCD FAINT} mode half of the time and {\tt 1-CCD
BRIGHT} the other half, and the GIS detectors in {\tt PH} mode. It was
also observed with the {\it ROSAT} PSPC on 1991 October 24. The
exposure times and count rates are summarized in Table~1. The data
were screened following standard procedures, as described in
\citet{eh98} and in \citet*{ehl96}, with the help of the {\tt
FTOOLS/XSELECT v.4.2} software package \citep*{bgp94,i94}. After
screening, we extracted spectra and light curves for each of the {\it
ASCA} detectors as well as a {\it ROSAT} PSPC spectrum. In the case of
the {\it ASCA} observations we opted to use {\tt BRIGHT} mode data so
that we could take advantage of the full exposure time. Although {\tt
BRIGHT} mode spectra represent a small compromise in spectral
resolution (because the echo effect and dark-frame error are not
corrected), their inclusion yields a significantly higher
signal-to-noise ratio (hereafter, $S/N$).

The galaxy MS~1718.6+4902 happened to fall within the field of view
of the {\it ROSAT} PSPC and the {\it ASCA} GIS during the observations
of Arp~102B. For the sake of completeness, we have extracted and
fitted the spectra of this source, as well. We present the results in
the appendix.

\begin{figure*}[t]
\begin{minipage}[t]{3.5in}
\centerline{\psfig{width=5in,figure=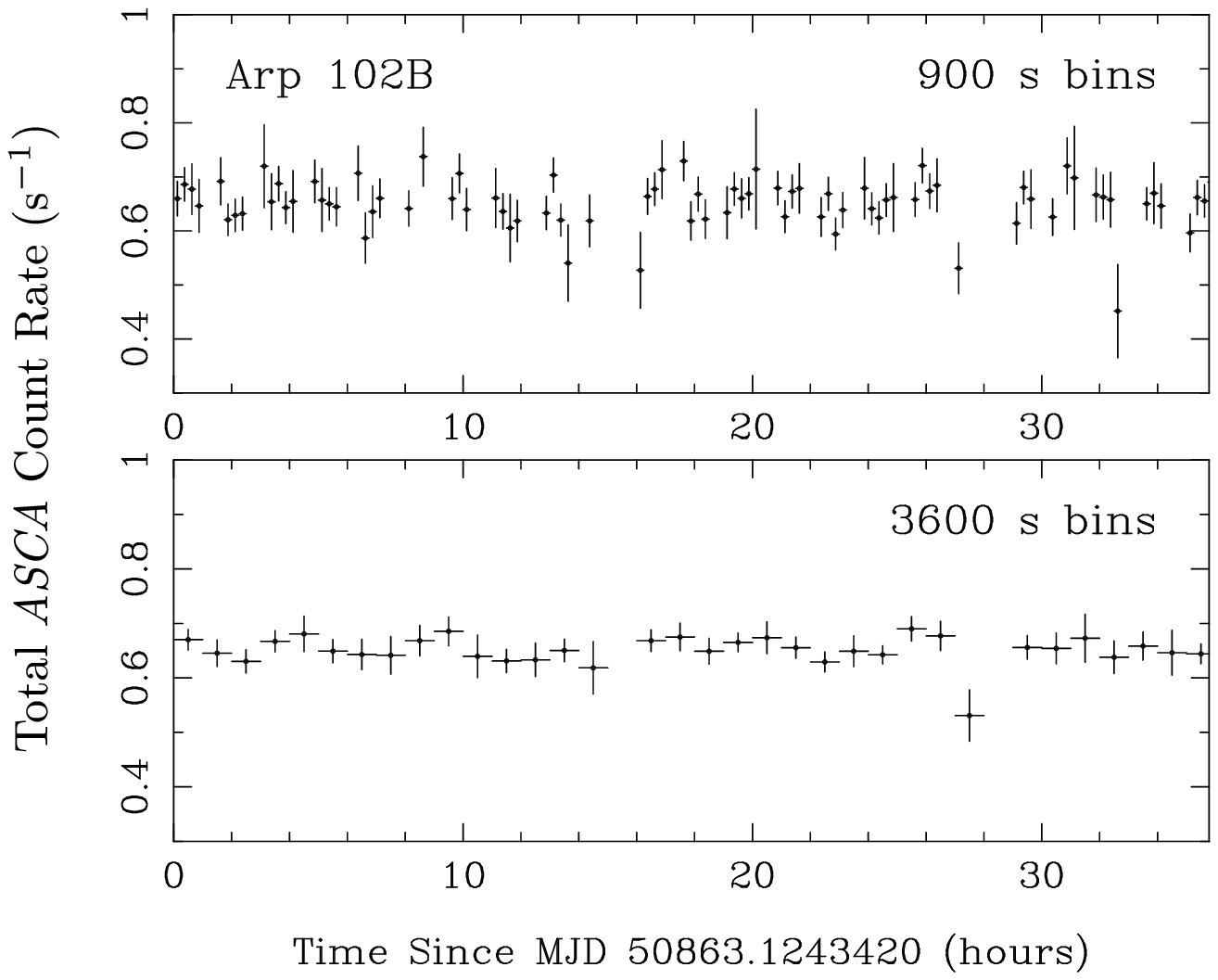}}
\figcaption{{\it ASCA} light curves of Arp~102B binned in (a) 15-minute bins
and (b) 1-hour bins. Each light curve shows the sum of counts detected by
all {\it ASCA} detectors.\label{fig_lc}}
\end{minipage}
\hskip 0.3truein
\begin{minipage}[t]{3.5in}
\centerline{\psfig{width=4.3in,figure=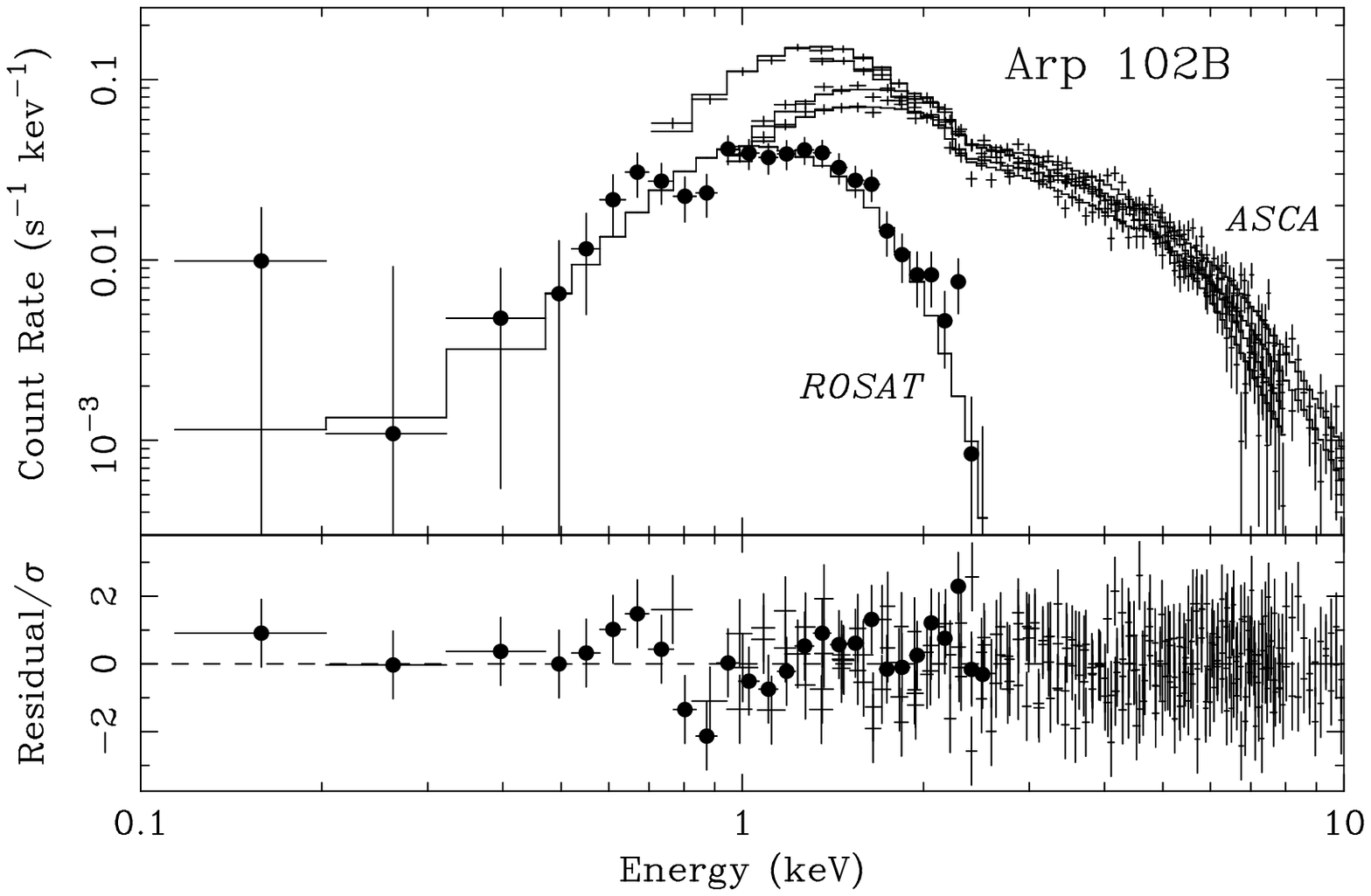}}
\figcaption{The upper panel shows spectra from each of the {\it ASCA} 
instruments and from the {\it ROSAT} PSPC with the best fitting model
superposed. The lower panel shows the residuals after subtraction of the model
from the observed spectra scaled by the error bar at each point. The model
plotted here is the simple power-law model that provides the best fit to the
{\it ASCA} spectra, with parameters as listed in Table~2. The same model has
been rescaled and superposed on the {\it ROSAT} spectrum.\label{fig_spec}}
\end{minipage}
\end{figure*}

\section{X-Ray Light Curves and Time Variability}

In Figure~\ref{fig_lc} we show two versions of the {\it ASCA} light
curve, one binned in 15-minute bins and the other binned in 1-hour
bins. Each light curve includes the sum of all counts from all four
detectors, to maximize the $S/N$.  The light curves show no obvious
variability on any time scale, from the shortest time scales sampled
to the length of the observation. To quantify the variability
properties of Arp~102B we carried out two types of tests:

\begin{enumerate}
\item
To search for rapid variability on time scales comparable to the light
curve sampling times we compared the r.m.s. fluctuations about the
mean to the average error bars in the light curve bins but found no
significant fluctuations.  We also computed the ``excess variance,''
$\sigma_{\rm rms}$, following \citet{n97a} and \citet{tur99},
obtaining $(1.0\pm0.1)\times 10^{-2}~{\rm s}^{-1}$ for the 15-minute
light curve and $(1.19\pm0.09)\times 10^{-2}~{\rm s}^{-1}$ for the
1-hour light curve. For a fair comparison with Seyfert galaxies we
also computed the excess variance for a light curve with 256~s bins,
which gave $(2.3\pm0.1)\times 10^{-2}~{\rm s}^{-1}$.  The above values
of the excess variance are similar to those found in Seyfert galaxies
of comparable luminosity ($L_{\rm 2-10~keV} \approx 3\times
10^{43}~{\rm erg~s^{-1}}$; see \S4) by \cite{n97a}.
\item
To search for long-term variations, on time scales comparable to the
length of the observation, we fitted the light curves with polynomials
looking for the lowest order that could provide an acceptable fit. We
found that the light curves are consistent with a constant.
\end{enumerate}

Finally, we searched for variability on much longer time scales by
comparing our measured flux with previous measurements. By fitting the
{\it ASCA} and {\it ROSAT} spectra (see next section and Table~2) we
find that the 1~keV flux density increased by a factor of 3 between
1991 and 1999. Similarly, an X-ray observation with the {\it Einstein}
IPC from 1980 \citep{b81} shows the 1~keV flux density to be 40\%
lower than the 1991 {\it ROSAT} measurement.

\section{Model Fits to the X-Ray Continuum and Determination of Absorption}

The observed spectra of counts {\it vs} energy channel were fitted
with continuum models to determine their shape and to measure the
column density of absorbing material along the line of
sight.\footnote{We carried out the fits with the help of the {\tt
XSPEC v.10} software package \citep{a96}. We used version 4 (1995
March 3) of the GIS response matrices and we generated SIS response
matrices appropriate for this particular observation using the {\sc
sisrmg} tool. To describe interstellar photoelectric absorption, we
adopted the cross-sections of \cite{mm83} and assumed that the heavy
elements have Solar abundances.} In our first attempt to fit the {\it
ASCA} spectra, with a simple model of a power law modified by
interstellar photoelectric absorption, we found a significant
discrepancy between the SIS$\,$1 spectrum and the spectra from the
other three detectors at low energies. To remedy this discrepancy,
which is the result of a progressive degradation of the SIS$\,$1
sensitivity at low energies \citep{yaqoob00}, we ignored all of the
SIS$\,$1 energy channels at energies below 1.2~keV. By taking this
measure, we were able to reconcile the results of fits to individual
spectra. We were also able to fit all spectra simultaneously with
common model parameters, allowing the normalization constants of
individual spectra to vary freely, to take account of uncertainties in
the absolute sensitivities of the four detectors. This simple
power-law model provides a good description of the observed spectrum,
with a relatively flat spectral index of $\Gamma=1.58\pm0.04$ and a
high absorbing column density of $N_{\rm H}=(2.8\pm0.3)\times
10^{21}~{\rm cm}^{-2}$ (error bars correspond to the 90\% confidence
level; see Table~2). The spectra, with models superposed are shown in
Figure~\ref{fig_spec}. This column is at least an order of magnitude
larger than the Galactic column. The Galactic reddening towards
Arp~102B is $E(B-V)=0.024$ \citep*{sfd98}, which implies and
equivalent \ion{H}{1} column of $N^{\rm Gal}_{\rm H} = 1.5\times
10^{20}~{\rm cm}^{-2}$, while measurements of the \ion{H}{1}~21~cm
line intensity towards Arp~102B yield $N^{\rm Gal}_{\rm H} = 2.3\times
10^{20}~{\rm cm}^{-2}$ \citep{s92}. The measured 2--10~keV flux
from this model is $F_{\rm 2-10~keV} = 1.2\times 10^{-11}~{\rm
erg~cm^{-2}~s^{-1}}$, which after correcting for absorption yields an
intrinsic luminosity of $L_{\rm 2-10~keV} = 3.1\times
10^{43}~{\rm erg~s^{-1}}$ (see Table~2).

\begin{figure*}[t]
\begin{minipage}[t]{3.5in}
\centerline{\psfig{width=4.3in,figure=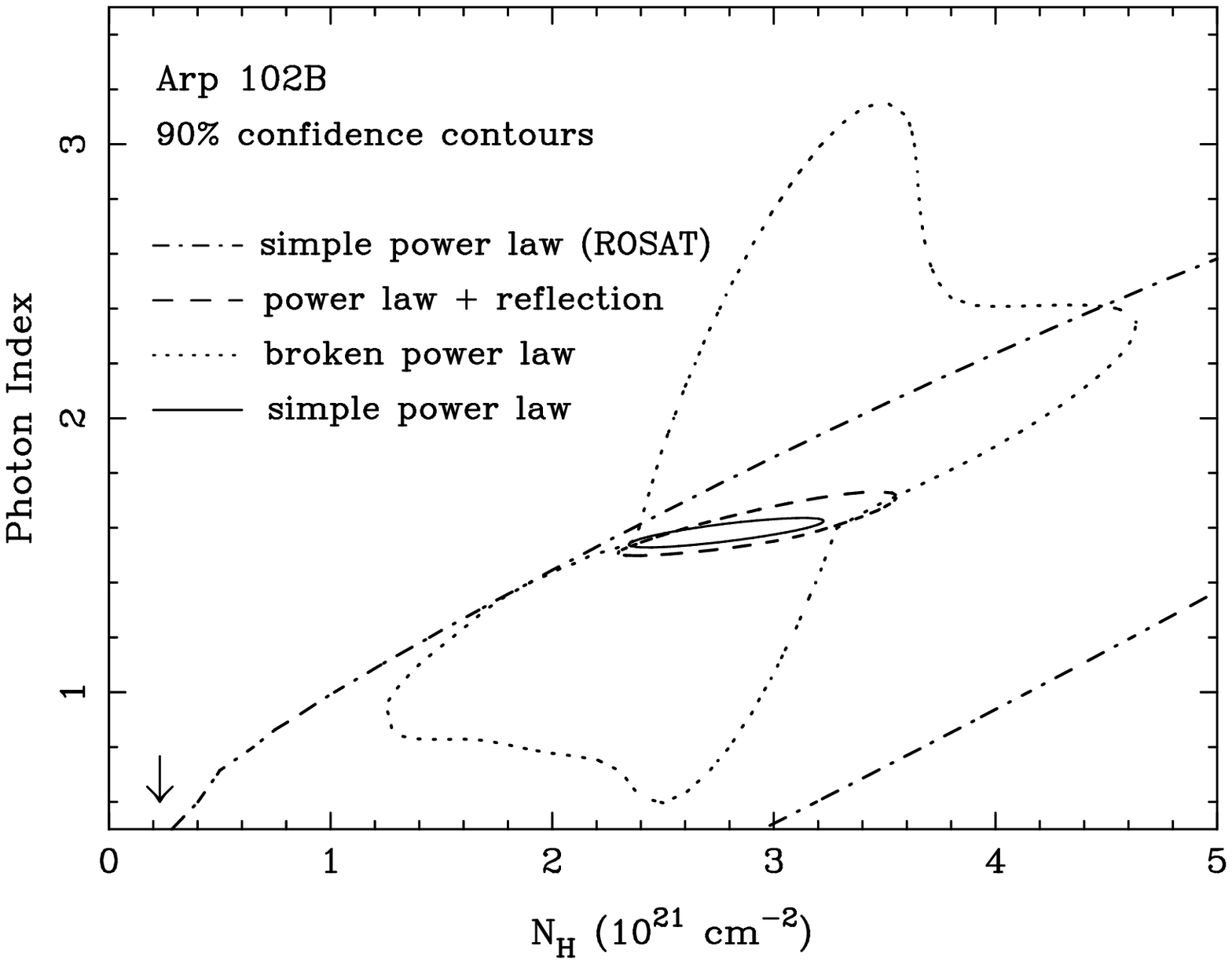}}
\figcaption{Contours showing the 90\% confidence region in the $\Gamma -
N_{\rm H}$ plane. Different line styles identify contours obtained
from different continuum models or different epochs, as follows: {\it
solid line}; simple power law fit to {\it ASCA} spectra; {\it dashed
line}, simple power law plus Compton ``reflection'' fit to {\it ASCA}
spectra; {\it dotted line}, broken power law fit to {\it ASCA}
spectra; {\it dot-dashed line}, simple power law fit to {\it ROSAT}
spectrum. In the case of the broken power law model the photon index
applies to the low-energy part of the spectrum, below the break. The
arrow in the lower left corner marks the Galactic column density in
the direction of Arp~102B.
\label{fig_cont}}
\end{minipage}
\hskip 0.3truein
\begin{minipage}[t]{3.5in}
\centerline{\psfig{width=4.3in,figure=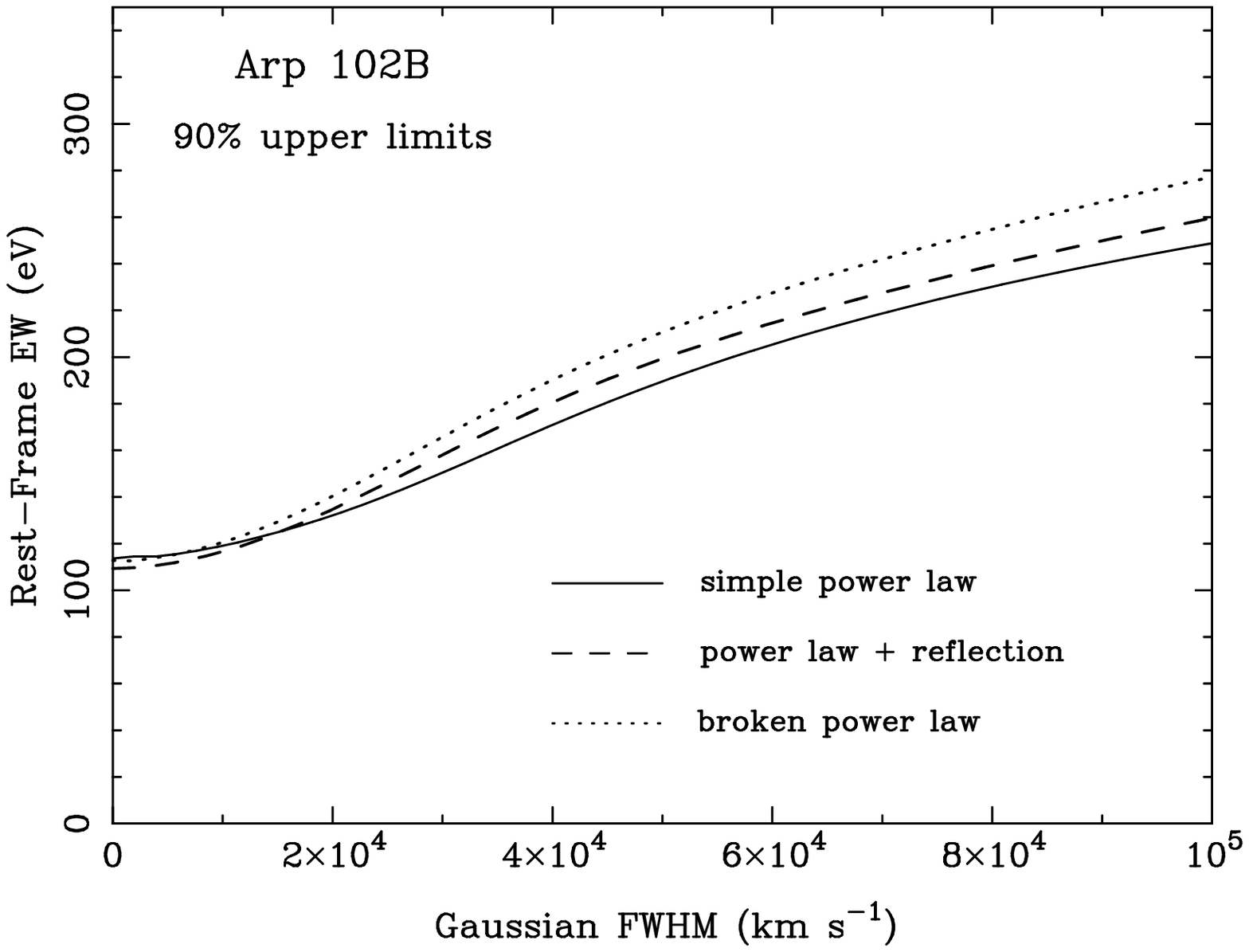}}
\figcaption{Upper limits to the equivalent width of the Fe~K$\alpha$ line
(at 90\% confidence) as a function of the assumed velocity width of
the line, for a Gaussian line profile. The three different line styles
refer to different assumed models for the continuum, as follows: {\it
solid line}, simple power law; {\it dashed line}, simple power law
plus Compton ``reflection;'' {\it dotted line}, broken power law.
\label{fig_line}}
\end{minipage}
\end{figure*}

Since the absorbing column density is central to our scientific goals,
we explored alternative continuum models in order to determine the
range of viable values. In particular, we fitted the continuum with a
broken power-law model to describe a possible soft excess (inspired by
Wo\'zniak et al. 1998\nocite{w98}) and with a Compton ``reflection''
model to describe a possible hard excess. In Table~2 we summarize the
parameters from the best fits. In fitting the broken power-law model
we constrained the break energy to lie between 1 and 4~keV, based on
the results of \citet{w98}. We find that the spectral indices above
and below the break converge to values that are equal to each other,
within uncertainties, indicating that there is no real break in the
spectrum. However, the additional freedom allowed by this model over
the simple power-law model results in a wider range of acceptable
values of the column density. In fitting the Compton ``reflection''
model we employed the {\sc pexrav} routine available within {\tt
XSPEC}, which computes the ``reflected'' spectrum as a function of the
inclination of the reprocessing slab by making use of the transfer
functions of \citet{mz95}. We assumed that the primary X-ray spectrum
incident on the reprocessor cuts off at 100~keV and that the heavy
elements in the reprocessor have Solar abundances. Thus the free
parameters of the model, in addition to the photon index and absorbing
column density, are the inclination angle of the reprocessing slab to
the line of sight, and the solid angle it subtends to the primary
X-ray source. We find that this model yields a very similar power-law
index and absorbing column density to the simple power-law model. The
inclination and solid angle of the reprocessing slab are virtually
unconstrained. This is hardly a surprise since the most pronounced
contribution of reprocessed X-rays is at energies higher than 10~keV,
which falls outside of the {\it ASCA} bandpass. The regions of the
$N_{\rm H}-\Gamma$ parameter space allowed by each of the above models
(at 90\% confidence) are shown in Figure~\ref{fig_cont}. We note that
although the region allowed by the broken power-law model is
considerably larger than those of the other two models, the minimum
allowed column density of $1.3\times~10^{21}~{\rm cm}^{-2}$ is still
well above the Galactic column density. In conclusion, we note that
neither of the alternatives to the simple power-law model provides a
better fit to the observed spectra (see the $\chi^2$ values in
Table~2).  Therefore, we prefer the simple power-law model because it
has the fewest free parameters. This conclusion is bolstered by the
fact that the spectra of of other BLRGs, observed up to 100~keV with
{\it RXTE} do not show strong reflection components \citep*{esm00}. In
fact, in half of the cases the 4--100~keV spectra are well-described
by a simple power law.

The {\it ROSAT} PSPC spectrum of Arp~102B, originally presented and
discussed by \citet{h97}, has a lower $S/N$ than the {\it ASCA}
spectra because of the heavy absorption and the soft bandpass of this
instrument.  Fitting this spectrum with an absorbed power-law model
yields only an upper limit to the photon index of $\Gamma < 2.2$ and a
wide range of allowed column densities, from $2\times 10^{20}$ to
$5\times 10^{21}~{\rm cm}^{-2}$ (at 90\% confidence). A complete list
of the best-fitting model parameters is given in Table~2. The {\it
ROSAT} PSPC spectrum is included in Figure~\ref{fig_spec}, along with
the {\it ASCA} spectra, with a rescaled version of the power-law model
that fits the {\it ASCA} spectra superposed. The constraints derived
from the {\it ROSAT} PSPC spectrum are plotted in
Figure~\ref{fig_cont} for comparison with the constraints derived from
the {\it ASCA} spectra. If we assume that the power-law photon index
during the {\it ROSAT} observation was at least 1.5, then the absorbing
column density must have been at least $2\times 10^{21}~{\rm cm}^{-2}$
at that time. Under the assumption that the spectral index of the {\it
ROSAT} PSPC spectrum is the same as that of the {\it ASCA} spectra, we
obtain an observed 0.5--2.0~keV flux of $F_{\rm 0.5-2.0~keV} =
1.3\times 10^{-12}~{\rm erg~cm^{-2}~s^{-1}}$. After correcting for
absorption, this flux yields an intrinsic luminosity of $L_{\rm
0.5-2.0~keV} = 3\times 10^{42}~{\rm erg~s^{-1}}$ (see Table~2).

Finally, we also searched for absorption edges from highly ionized
metals, such as \ion{O}{7} and \ion{O}{8}, at 0.74 and 0.87 keV
respectively, since measurements of, or limits on, the optical depths
of these edges constrain the properties of the absorber. These edges
occur at energies where the degraded SIS$\,$1 sensitivity leads to
discrepancies between the continuum model fits to the SIS$\,$1 and
SIS$\,$0 spectra, and specifically to the inferred equivalent hydrogen
column density. To bypass this problem we used the SIS$\,$0 and
SIS$\,$1 spectra down to an energy of 0.6~keV and allowed the column
densities fitted to the two spectra to be independent parameters. This
way we arrive at a parametric description of the continuum, which is
not physically meaningful, but still allows us to measure discrete
absorption features relative to the continuum level in both the
SIS$\,$0 and SIS$\,$1 spectra. By adding an edge to our spectral model
and scanning a range of edge rest energies between 0.7 and 0.9~keV we
were able to determine the following optical depth limits at the 90\%
confidence level: $\tau < 0.28$ for the \ion{O}{7} edge and $\tau <
0.25$ for the \ion{O}{8} edge.

\section{The Fe K$\alpha$ Line: Too Weak to Be Measurable}

Since Arp~102B is notorious for its disk-like, double-peaked, optical
emission lines and since the X-ray spectra of Seyfert galaxies are
well-known to include disk-like Fe~K$\alpha$ lines (e.g., Nandra et
al. 1997b\nocite{n97b}), we have looked for the Fe~K$\alpha$ line in
the X-ray spectrum of Arp~102B. Unfortunately, we were not able to
detect it, although we were able to place interesting limits on its
$EW$. To search for the line we fitted a model to the continuum
excluding from the fit the energy range between 5 and 7~keV, where the
Fe~K$\alpha$ line is expected to be (assuming a rest energy of
6.4~keV). We then froze the continuum model parameters, included a
Gaussian line profile in the model, and added back to the spectra the
energy channels between 5 and 7~keV. To determine upper limits to the
line $EW$ we scanned the line flux--velocity dispersion parameter
plane looking for regions where the model fits was acceptable, keeping
the rest energy of the line fixed. Since the continuum was held fixed
throughout this procedure, the resulting upper limits on the line flux
can be converted to upper limits on its $EW$ via a simple scaling. To
check whether the assumed continuum model affects the results we
repeated the exercise for all of the continuum models discussed in
\S4. Our findings are shown graphically in Figure~\ref{fig_line},
where we plot upper 90\% confidence limits to the {\it rest} $EW$ of
the line against its assumed full width at half maximum (hereafter,
FWHM). The assumed continuum model has a very small effect on the
derived limits, which range from 120~eV for an unresolved line to
about 200~eV for a line of FWHM comparable to that of Seyfert galaxy
Fe~K$\alpha$ lines (60,000~km~s$^{-1}$; Nandra et al.
1997b\nocite{n97b}). This limit is below the typical $EW$s measured
for Seyfert galaxy Fe~K$\alpha$ lines, as we discuss further in \S7.1.

\begin{figure*}[t]
\begin{minipage}[t]{3.5in}
\centerline{\psfig{width=4.3in,figure=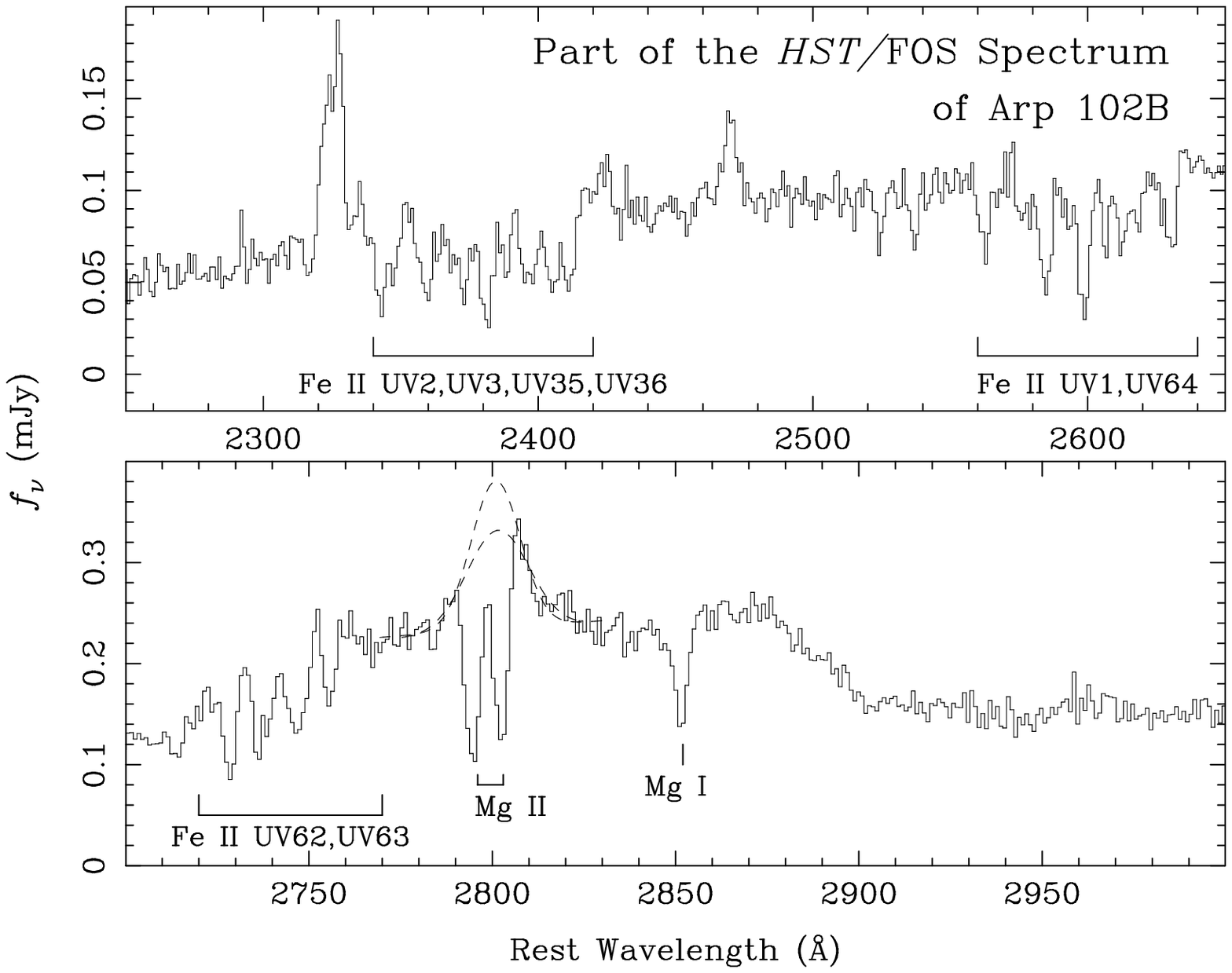}}
\figcaption{A segment of the UV spectrum of Arp~102B observed with the
{\it HST} \citep{h96}, showing the the plethora of \ion{Fe}{2}
absorption lines as well as the \ion{Mg}{1} and \ion{Mg}{2} absorption
lines. The broad \ion{Mg}{2} {\it emission} line and a number of
narrow emission lines are also evident. The dashed lines show two
extreme fits to the peak of the \ion{Mg}{2} line, relative to which
the absorption lines are measured: they illustrate the uncertainty
resulting from the process of defining an effective continuum.
\label{fig_hst}}
\end{minipage}
\hskip 0.3truein
\begin{minipage}[t]{3.5in}
\centerline{\psfig{width=4.3in,figure=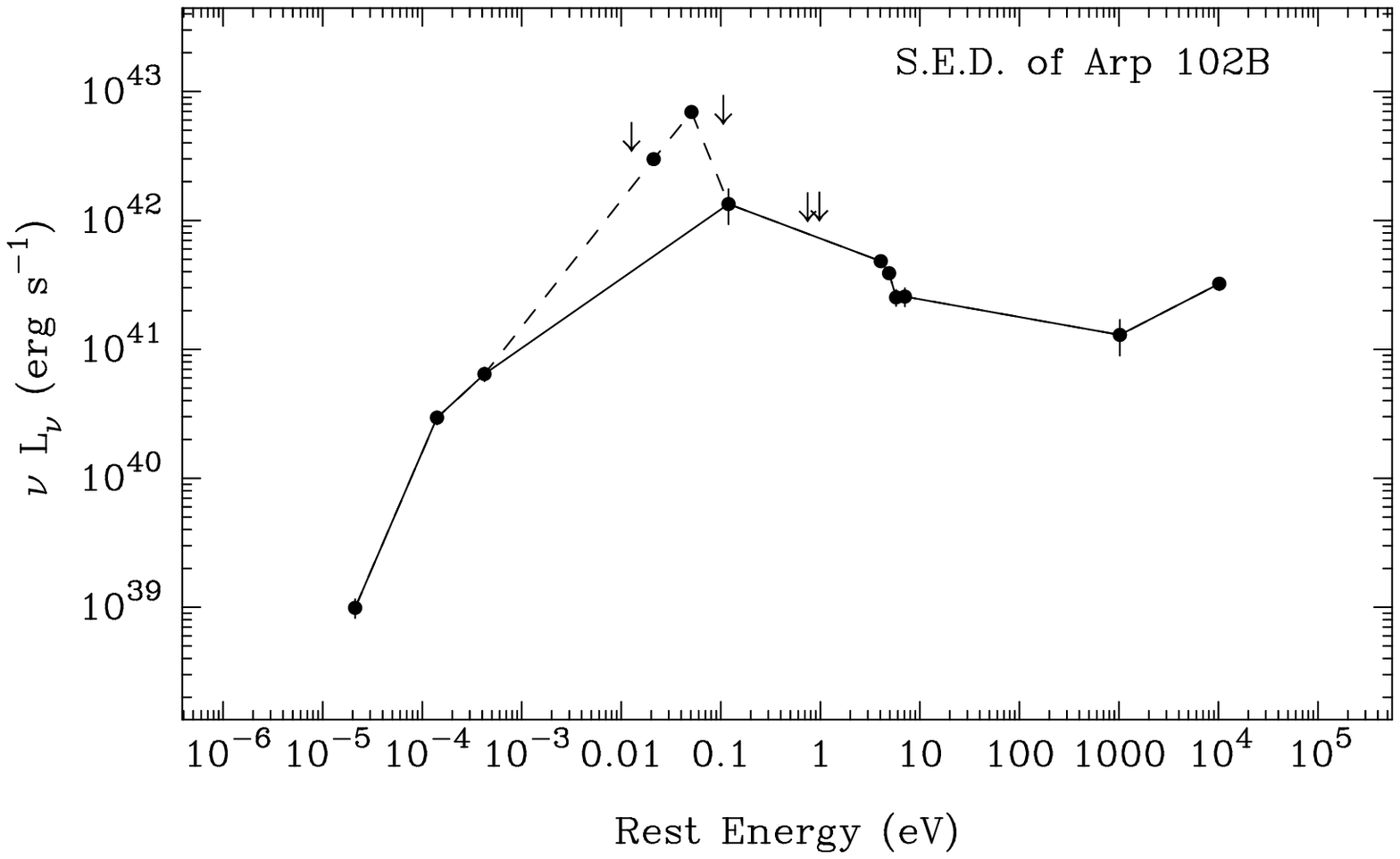}}
\figcaption{The data points show the observed spectral energy
distribution (SED) of Arp~102B, which was used in our photoionization
calculations. The X-ray data points at 1 and 10~keV are from the {\it
ASCA} measurements presented here. The UV data points (between 1 and
10 eV) are from the {\it HST} spectrum \citep{h96}. The arrows near
1~eV are upper limits from the J- and H-band spectra of
\citet{sfbw96}.  All other data points are from the compilation of
\citet{ch89}, where references to the original papers are given. The
arrows near 0.01 and 0.1~eV denote upper limits from IRAS observations
at 12 and 100~$\mu$m. The points connected by a solid line define the
SED {\it without} a far-IR bump used in one of our two photoionization
calculations. For the other photoionization calculation we used all of
the available measurements to define an SED {\it with} a far-IR bump.
The far-IR bump is represented by the IRAS points, which are connected
by a dashed line. A more extensive discussion of the choice of SEDs is
given in \S6.2 of the text.
\label{fig_sed}}
\end{minipage}
\end{figure*}

\section{Evaluation of Simple Models for the UV (and X-Ray) Absorber}

Using the column density measured from the X-ray spectrum of Arp~102B
and the properties of its UV absorption lines, we constructed simple
photoionization models for the absorbing gas. The goal was to
constrain the physical conditions of the absorber and its location
relative to the continuum and broad-emission line sources, and deduce
its role in the overall accretion flow.  Previous attempts to model
metastable \ion{Fe}{2} absorbers in luminous quasars include those of
\citet{wcp95} who studied Q~0059--2735 and of \citet{dek01,dek02} who
studied a number of quasars found in the FIRST survey.  The above
studies employed high-resolution spectra, which made possible the
direct determination of the column densities of ionic species of
interest and a diagnosis of the physical conditions in the
absorber. Because of the low resolution of our UV spectra, we make a
number of simplifying assumptions in order to construct a simple model
and then compare its predictions to the data. This exercise tests a
specific hypothesis, rather than eliminating alternatives. In later
sections we discuss possible observations that can lead to further
tests and refinements of this model.

\subsection{Limitations of the UV Spectra and Measurement of 
Absorption Lines}

A segment of the UV spectrum of Arp~102B in the vicinity of the
\ion{Mg}{2} line, taken with the {\it Hubble Space Telescope (HST)} is
shown in Figure~\ref{fig_hst}, for reference.  The full
(1200-3200~\AA) UV spectrum along with the details of the observations
can be found in \citet{h96}. A very serious limitation of the UV
spectrum is its modest spectral resolution, which does not allow us to
resolve the profiles of the absorption lines fully. As a result we
cannot be certain whether the UV absorption lines are saturated, we
cannot determine the kinematic structure of the absorbing gas, and we
cannot unambiguously ascertain whether the absorber fully covers the
continuum and broad-emission line sources \citep[for a more extensive
discussion of partial coverage see][]{bs97,ham97,gang99}.  We can,
nevertheless, draw indirect inferences about the above from the widths
of the UV absorption lines, most importantly the lines of \ion{Mg}{2}
doublet. The observable quantities we rely on, therefore, are the
$EW$s of the UV absorption lines, their full widths at half minimum
(FWHMin), and the equivalent hydrogen column density measured from the
X-ray spectrum.

We measured the rest $EW$s and FWHMin of the UV absorption lines
relative to an effective continuum, defined by the combination of the
true continuum and the broad emission lines on which they are
superposed. The effective continuum was determined by fitting a linear
combination of a low-order polynomial and one or two Gaussians to the
region around the absorption lines of interest.  We list the measured
quantities in Table~3. A source of uncertainty is the definition of
the effective continuum since the absorption lines are typically
superposed on emission lines whose profiles are severely distorted by
the absorption lines themselves. This is the dominant source of
uncertainty in the measurements given in Table~3. We have estimated
the uncertainty in $EW$s and FWHMin by experimenting with a variety of
continuum fits and noting their effect on the measured quantities.  To
illustrate the magnitude of these uncertainties we plot two extreme
continuum fits to the peak of the \ion{Mg}{2} line in
Figure~\ref{fig_hst}. These extreme fits set the extreme values of the
$EW$s, while the best estimate is taken to be the average of the two
extremes. In the case of Ly$\alpha$, there is an additional source of
uncertainty, which we cannot quantify: absorption by the interstellar
medium in the host galaxy of Arp~120B. To deal with this issue we will
treat the Ly$\alpha$ $EW$ with caution in our later analysis.

The properties of the \llion{Mg}{2}{27896,2803} doublet allow us to
place some constraints on the coverage fraction of the background
source(s) by the absorbing gas. The FWHMin we measure corresponds to a
velocity of approximately 580~km~s$^{-1}$, which is much higher than
the instrumental resolution of 210~km~s$^{-1}$. However, it is very
unlikely that this broadening is due to a Gaussian internal velocity
spectrum in a single parcel of absorbing gas for the following reason.
We have simulated\footnote{The simulation consists of synthesizing a
theoretical profile for an assumed column density and kinematic
broadening parameter (i.e., Gaussian velocity dispersion) and then
convolving it with the instrumental line profile. The theoretical line
profile is synthesized following the formalism of \citet{hum79}. The
instrumental line profile is taken to be a Gaussian whose width
depends on the grating-detector combination \citep{keys95}.} observed
\ion{Mg}{2} line profiles for a range of column densities and
broadening parameters (i.e., Gaussian velocity dispersions, denoted by
$b$) and found that the lines should be fully blended if the
broadening parameter was indeed as high 580~km~s$^{-1}$, which is not
observed. If, instead, we assume that the column density is so high
that the lines are saturated and the kinematic broadening is small
($b<200\kms$), we find that the lines are resolved and their depths
are approximately equal, in very good agreement with
observations. This is because the theoretical profiles of the
saturated lines are approximately square so that their widths at half
minimum can be large, without any blending of the wings.  Further
support for saturation of the \ion{Mg}{2} doublet is provided by the
observation that the two lines have nearly equal strengths even though
their oscillator strengths have a ratio of 2:1. We therefore adopt the
latter picture as our working hypothesis, noting the following
important caveat: saturation is not necessarily the only way of
reproducing the properties of the \ion{Mg}{2} profiles. Other
possibilities include a highly-ordered velocity field (as in an
accelerating outflow, for example) or a clumpy absorber which gives
rise to multiple ``components'' in the line profiles. Since neither of
these possibilities can be constrained with the available data, we
cannot explore them further.

An important consequence of the saturated \ion{Mg}{2} lines is that
the absorber must cover the background sources (continuum source
and/or broad-emission line region) only partly; if it did not, the
troughs of the absorption lines should have been nearly black, (i.e.,
they should have reached close to zero intensity, since their widths
are just below the resolution limit of the instrument). From the
observed depth of the lines and by comparison with our simulated
profiles, we deduce a coverage fraction of $f_{\rm C}\approx 0.5$,
which we use to correct the measured $EW$ of the \ion{Mg}{2} lines
(see Table~3). Carrying out the same exercise for other absorption
lines (Ly$\alpha$, \ion{Si}{3}, and \ion{Mg}{1}) we derive comparable
coverage fractions, which we also list in Table~3. Finally, we also
assume a coverage fraction of 0.6 for the weaker lines for the sake of
consistency. As we show later, this assumption does not affect our
results or conclusions. These coverage fractions should be regarded
with caution because they are inferred indirectly and only after
making the assumptions that we do.  An alternative possibility, for
example, is that the lines originate in an outflow whose ordered
velocity field leads to square absorption-line profiles.  Therefore,
in the analysis that follows, we will explore both the case of finite
and negligible partial coverage.

\subsection{Model Assumptions and Methodology}

Our main observational constraints comprise the total column density
of hydrogen, determined from the X-ray spectrum, and the $EW$s of the
UV absorption lines. An additional constraint is provided by the
relative strengths of the \ion{Fe}{2} absorption lines from the ground
state and metastable states. More specifically, the lines of the UV1
and UV64 multiplets, which fall in a ``clean'' part of the UV
continuum, around a rest wavelength of 2600~\AA, can be used. Thus, in
order to construct a model for the absorber we make the following
assumptions:
\begin{itemize}
\item
The X-ray and UV absorption arise in the same medium, therefore we
seek a model that can satisfy all the observational constraints 
simultaneously, as far as possible.
\item
The absorber consists of a uniform slab of gas, which is photoionized
by the continuum from the AGN central engine. This slab most likely
has a bulk outflow velocity of about 200~km~s$^{-1}$, as noted by
\citet{h96}, but its internal velocity dispersion is assumed to be
small ($< 200~{\rm km~s}^{-1}$; see \S6.1). The equivalent hydrogen
column density of this slab is $N_{\rm H}=2.8\times 10^{21}~{\rm
cm}^{-2}$, as measured from the {\it ASCA} spectra and its metallicity
and abundance pattern follow those of the Sun.
\item
The spectral energy distribution (SED) of the ionizing continuum is
the same as the observed SED, which we plot in Figure~\ref{fig_sed}.
We assume that the shape of the SED between 1 and 10~keV continues up
to 100~keV where it cuts off exponentially.  An important feature of
this SED is that it has no ``UV bump.'' As such it closely resembles
the SEDs of low-luminosity AGNs \cite[see][]{Ho99}. The far-IR
measurements of this SED come from IRAS observations, which were taken
through a very large aperture and may not represent the true far-IR
flux of the nucleus. Therefore, we repeated our calculations for two
different assumed SEDs, with and without a ``far-IR bump'' (see
illustration in Figure~\ref{fig_sed}).

\end{itemize}

Under the above assumptions we computed the ionization structure of
the absorbing slab using the photoionization code Cloudy
\citep[versions 94.00 and 95.04$\beta$3;][]{cloudy} for a wide range
of values of the ionization parameter and hydrogen number density ($U$
and $n$, respectively). In particular, our calculations span the range
$\log U = -5.0~{\rm to}~+2.0$ in steps of 0.5 and $\log n = 2~{\rm
to}~15$ in steps of 1 ($n$ is measured in cm$^{-3}$). The code
includes a detailed model of the \ion{Fe}{2} ion \citep{verner} which
yields the optical depths at line center of the \ion{Fe}{2}
transitions of interest. For each cell in the $\log U - \log n$
parameter space we obtain the column density of the ions of interest,
from which we compute the $EW$s of absorption lines for an assumed
broadening parameter. These are the quantities that we can compare
directly with observations. By comparing the observed and predicted
$EW$s of each observed absorption line we obtain a set of constraints
on $\log U$ and $\log n$, as we describe further in the next section.
The lines used in this comparison are those listed in Table~3. In the
second stage of our comparison of models with data, we derive
additional constraints by using the relative strengths of the lines in
the \ion{Fe}{2} UV1 and UV64 multiplets and the shape of the soft
X-ray spectrum.

\begin{figure*}[t]
\begin{minipage}[t]{3.5in}
\centerline{\psfig{width=4.3in,figure=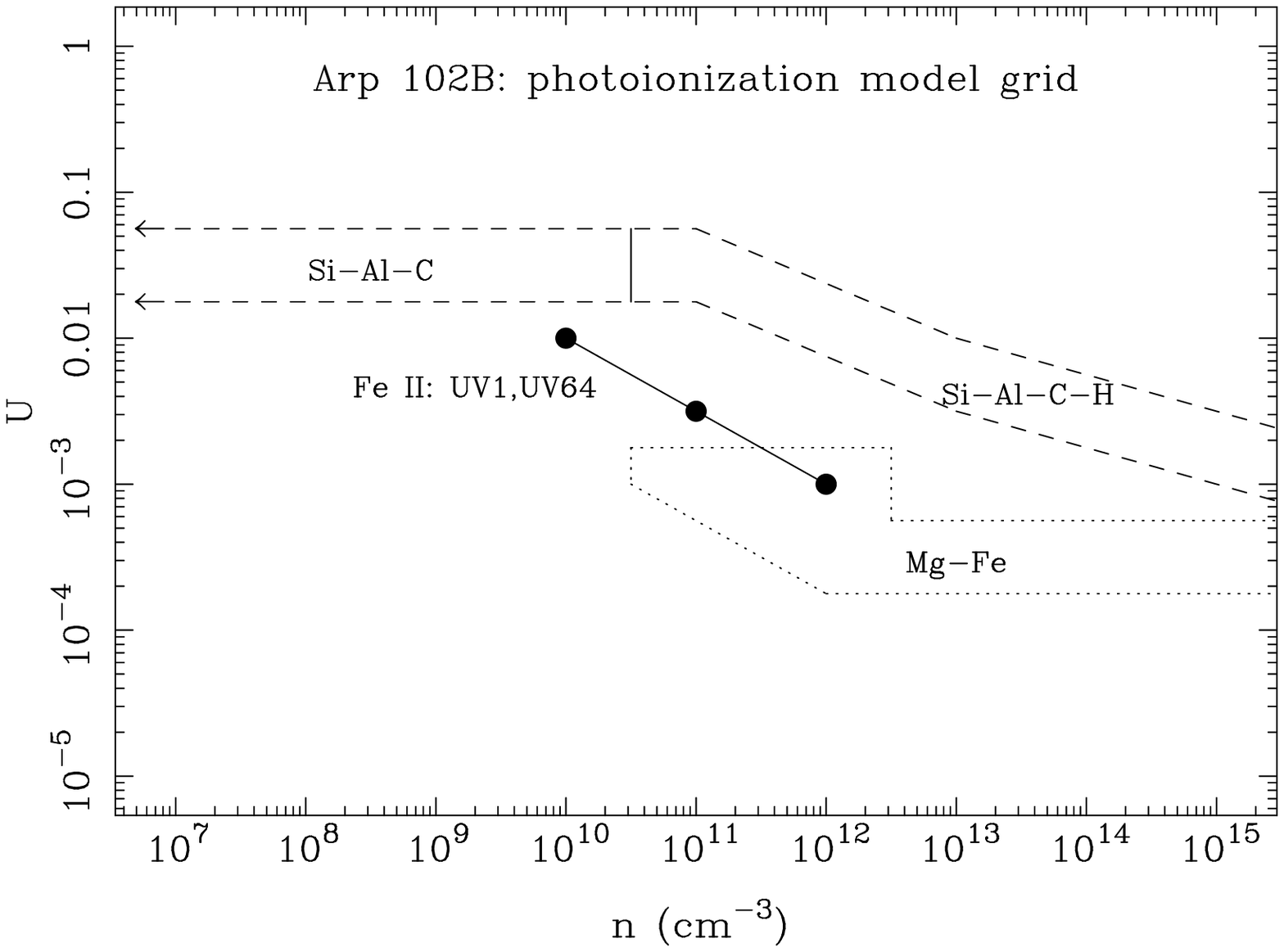}}
\figcaption{Constraints on the ionization parameter and hydrogen density
from the observed absorption line $EW$s. This diagram shows a segment
of the $\log U - \log n$ parameter space spanned by our
photoionization calculations described in \S6.2 of the text. The
dotted line shows the region allowed by the $EW$s of the Mg, and Fe
lines.  The dashed line shows the region allowed by the $EW$s of the
Si, Al, and C lines (no density constraints; labeled as Si-Al-C). If
the $EW$ of Ly$\alpha$ is used as an additional constraint, only the
part of this region on the right of the solid vertical bar is allowed
($\log n \geq 11$; labeled as Si-Al-C-H). The filled circles
connected by a solid line mark the grid cells for which the predicted
spectrum of the \ion{Fe}{2} UV1 and UV64 multiplets is in reasonable
agreement with the data (see Figure~\ref{fig_fe}). The SEDs with and
without a far-IR bump produce virtually identical results.
\label{fig_grid}}
\end{minipage}
\hskip 0.3truein
\begin{minipage}[t]{3.5in}
\centerline{\psfig{width=4.in,figure=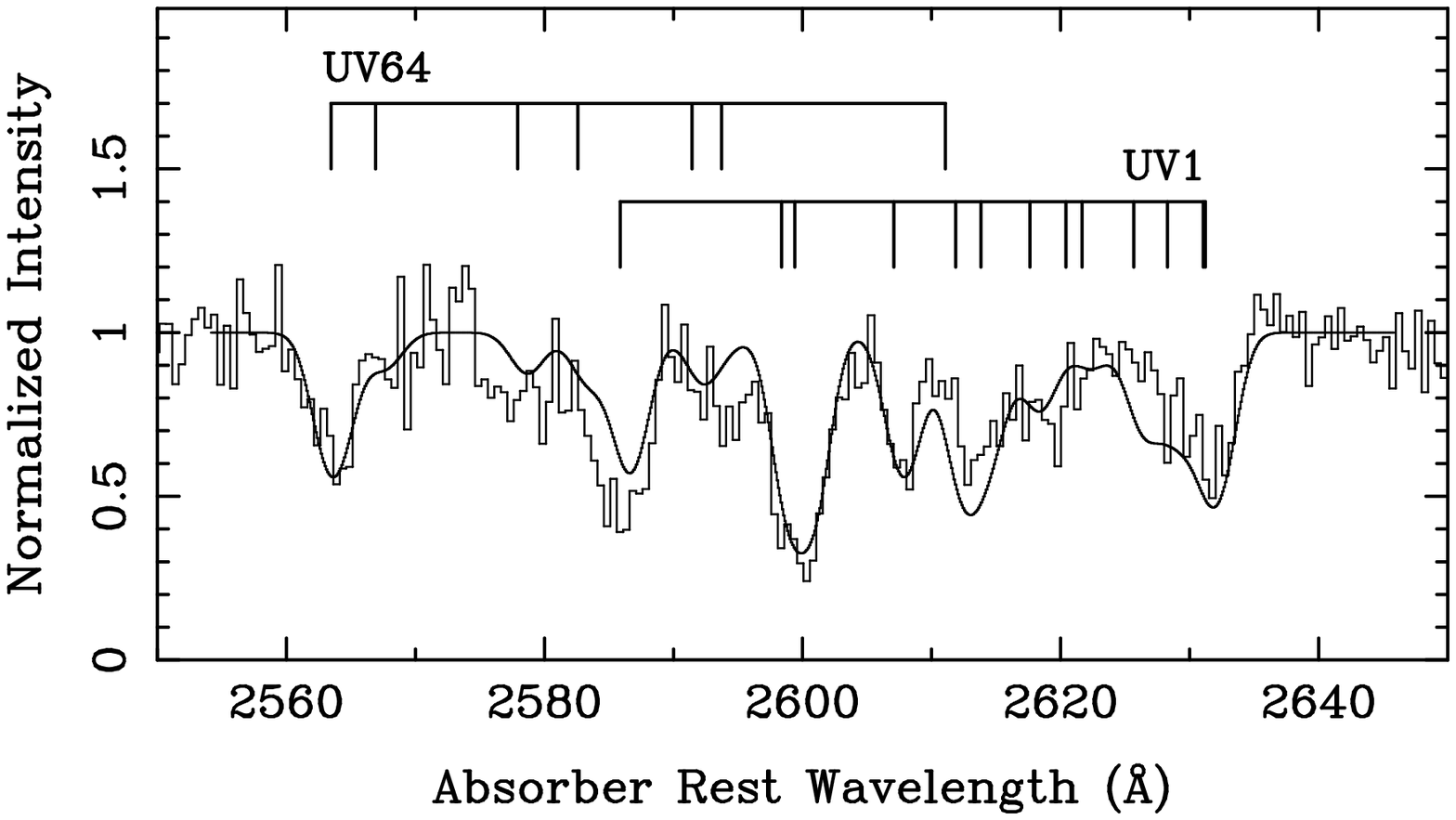}}
\figcaption{Comparison of the predicted spectrum of the \ion{Fe}{2} UV1
and UV64 multiplets (smooth line), with the normalized spectrum of
Arp~102B. The theoretical spectrum corresponds to a model with $\log
U=-3$ and $\log n = 12$, with broadening parameter $b=180\kms$ and a
coverage fraction $f_{\rm C}=0.7$.  The locations of the lines from each
multiplet are identified.  The wavelength scale refers to the rest
frame of the absorber, which is blueshifted by 150\kms\ relative to the
frame of the narrow emission lines of Arp~102B. 
\label{fig_fe}}
\end{minipage}
\end{figure*}

\subsection{Results and Checks for Self-Consistency}

Our results are shown graphically in Figure~\ref{fig_grid}, which
depicts part of the $\log U-\log n$ parameter space covered by our
photoionization calculations. The two different assumed SEDs give
virtually identical results, therefore we describe and discuss the
results below without regard to the SED.  The dotted line in this
figure shows the region in which the $EW$s of the \ion{Mg}{1},
\ion{Mg}{2}, and \ion{Fe}{2} lines predicted by the models agree with
the measured $EW$. The lower limit to the density is set by the $EW$
of the \lion{Mg}{1}{2853} line. Thus the properties of the absorber
are constrained by these lines as follows: $-3.5 < \log U < -2.5$ and
$\log n \geq 11$. The models that reproduce the $EW$s of the Mg-Fe
family of lines cannot reproduce the $EW$s of all the other lines of
Table~3 (the Si-Al-C-H family) for a single broadening parameter. The
latter family of lines requires similar densities to the former, but a
somewhat higher ionization parameter, namely $-3.0 < \log U < -1.5$
and $\log n \geq 11$, as shown by the dashed line in
Figure~\ref{fig_grid}.  If we do not consider the $EW$ of Ly$\alpha$
(see discussion in \S6.1) as one of our constraints, the Si, Al, and C
do not set any limits on the density of the absorber. Inclusion of
Ly$\alpha$, however, does set a lower limit on the density of $\log n
\geq 11$ (solid vertical bar in Figure~\ref{fig_grid}) as follows: for
lower densities, one can find models that reproduce the $EW$s of the
Si, Al, and C lines simultaneously but fail to reproduce the $EW$ of
Ly$\alpha$. This set of model parameters predicts a $EW$ for the Mg
and Fe lines that is too low compared to observations.

It is noteworthy that the above results are insensitive to the
assumption of partial coverage.  According to the models, all of the
observed absorption lines are on the flat part of the curve of growth,
therefore their theoretical $EW$s depend mostly on the broadening
parameter and are relatively insensitive to the ionic column
density. If we take the measured $EW$s at face value, without applying
a correction for partial coverage, we find that the lines in the Mg-Fe
family require $b=80-100\kms$ and the lines in the Si-Al-C-H family
require $b=90-130\kms$. If, on the other hand, we adopt the coverage
fractions listed in Table~3, we find that the lines in the Mg-Fe
family require $b=140-190\kms$ and the lines in the Si-Al-C-H family
require $b=100-160\kms$.

As a consistency test we compared the relative strengths of the lines
in the \ion{Fe}{2} UV1 and UV64 multiplets with the model
predictions. These are the only useful \ion{Fe}{2} multiplets for this
test because the fall in ``clean'' part of the continuum, which is
free of emission lines. Since the lines are severely blended, we used
the optical depths predicted by the models, assumed a value for the
broadening parameter and synthesized the expected spectrum, which we
compared to the data. The adjustable parameters in this comparison
were the broadening parameter and the coverage fraction of the
background source by the absorber. We searched the parameter region
defined by $\log U = -2.0$ to $-3.5$ and $\log n = 10$ to 15 and found
reasonable, although not perfect, agreement for the following
parameter sets: $(\log U,\,\log n)=(-2.0,\,10),\, (-2.5,\,11),\,
(-3.0,\,12)$ and coverage fractions between 0.7 and 0.8. These
parameter sets are indicated in Figure~\ref{fig_grid} with large
filled circles connected by a solid line. The best agreement between
the observed and predicted spectrum is achieved for $(\log U,\,\log
n)=(-3.0,\,12)$, $b=180\kms$, and $f_{\rm C}=0.7$. The predicted
spectrum resulting from this model is compared to the observed
spectrum in Figure~\ref{fig_fe}, where the lines from the \ion{Fe}{2}
UV1 and UV64 multiplets are marked.  We note that in
Figure~\ref{fig_fe}, the relative strengths of the strongest lines
from each multiplet (\lam2600 from UV1 and \lam2563 from UV64) agree
fairly well with the data, but the relative strengths of lines within
a multiplet do not.\footnote{This could be a consequence of one or
more of the following factors: (a) Blending of some of the \ion{Fe}{2}
lines of Arp~102B with Galactic \ion{Fe}{2} lines. As shown in
\citet{h96}, some weak \ion{Fe}{2}~UV1 lines at $z=0$ overlap with
some of the \ion{Fe}{2}~UV64 lines at $z=0.0244$. This is a minor
effect, however, and cannot explain the discrepancy fully. (b) Coarse
temperature or density sampling in our model grid. These multiplets
could conceivably provide much finer constraints on the temperature
and density, but the necessary computations are prohibitively
expensive. (c) The \ion{Fe}{2} oscillator strengths used by Cloudy,
which are taken from the Iron Project \citep{nahar95} may not be quite
correct.  \citet{dek01} compared measured and calculated \ion{Fe}{2}
oscillator strengths from various sources and found significant
outliers among the Iron Project values. An inspection of their Table~2
shows a marked correspondence between the poorly-fitting lines in our
Figure~\ref{fig_fe} and the discrepant oscillator strengths.} At $\log
n < 10$ or $\log U < -3.5$, the optical depths of these lines are so
large that they blend together. At $\log n > 12$ the lines become
rather weak due to collisional ionization of \ion{Fe}{2} to
\ion{Fe}{3} and their predicted relative strengths are in very poor
agreement with the observations.

Additional consistency tests are afforded by the X-ray spectrum and in
particular, by the observed equivalent neutral hydrogen column density
and the absence of absorption edges from highly-ionized metals. The
models that best explain the UV absorption lines predict that the
elements that contribute to the soft X-ray absorption
\citep[0.1--1~keV; see][]{mm83}, namely C, N, and O, retain their
K-shell electrons and can act as effective absorbers. For example, at
$\log U = -3.0$ the dominant ionization stages of these elements are
\ion{C}{1}, \ion{C}{2}, \ion{N}{1}, \ion{N}{2}, \ion{O}{2}, and
\ion{O}{3}.  At $\log U = -1.5$, however, an appreciable fraction of C
and N atoms become fully ionized, which reduces the soft X-ray opacity
and an appreciable fraction of the O atoms are in the form of
\ion{O}{7} and \ion{O}{8}. This results in strong absorption edges at
0.74 and 0.87~keV respectively. Therefore, the X-ray spectrum
constrains the ionization parameter to values of $\log U \leq -2.0$,
where soft X-ray absorption by metals is appreciable and the optical
depths of \ion{O}{7} and \ion{O}{8} K-shell edges are below our
observed upper limits (see \S4).  At the lowest X-ray energies of the
{\it ROSAT} spectrum (0.1--0.2 keV), the opacity is dominated by
photoelectric absorption by He, which at the densities of interest, is
mostly in the form of \ion{He}{2} and \ion{He}{3} at $\log U \geq
-2.5$ and mostly in the form of \ion{He}{1} and \ion{He}{2} at $\log U
\leq -3.0$. This implies that the soft X-ray flux should start to
recover from absorption at such low energies.  Unfortunately, the poor
$S/N$ of the {\it ROSAT} spectrum at these low energies does not allow
us to check this prediction, although the shape of the spectrum is
suggestive of an upturn at 0.1--0.2~keV (see Figure~\ref{fig_spec}).

\section{Discussion and Conclusions}

\subsection{The X-Ray Spectral Properties of Arp~102B in Context}

The X-ray properties of Arp~102B reinforce previously known trends of
systematic differences between Seyfert galaxies and BLRGs.  As we have
already remarked, simple power-law spectra, with neither a soft nor a
hard excess are fairly common among BLRGs observed with {\it ASCA},
{\it RXTE}, {\it BeppoSAX}, and {\it Ginga} (Wo\'zniak et al. 1998;
Sambruna et al. 1999; Hasenkopf, Sambruna, \& Eracleous 2002;
Eracleous et al. 2000; Grandi, Urry, \& Maraschi 2002; Zdziarski et
al. 1995)\nocite{w98,sem99,hse02,esm00,gum02,z95}.  Seyferts, on the
other hand, often show a soft and/or a hard excess in their spectra.
The heavy absorption by ``cold'' (i.e. neutral) matter found in
Arp~102B is not uncommon in radio-loud AGNs, in contrast to Seyfert~1
galaxies whose spectra often show the signature of a ``warm'' (i.e.,
ionized) absorber \citep{r97,g98,k00}.  \citet{sem99} find such heavy
absorption in 1/3 of the the BLRGs and radio-loud quasars in their
collection, with columns of $10^{21}~{\rm cm}^{-2}$, or higher (see
their Figure~6).

The relatively flat spectral index of Arp~102B, although not typical
of luminous BLRGs (e.g., 3C~390.3, 3C~111, 3C~120, 3C~382) is
characteristic of a subclass of radio-loud AGNs, the weak-line radio
galaxies (hereafter WLRGs; see Sambruna et al.  1999\nocite{sem99} for
a summary of their X-ray properties). These are radio galaxies
distinguished by the low luminosity of their [\ion{O}{3}]$\;$\lam5007
emission lines \citep{t98,l94}. The combination of spectral index and
X-ray luminosity of Arp~102B puts it well within the region occupied
by WLRGs in Figure~2d of \citet{sem99}. Moreover, its
[\ion{O}{3}]$\;$\lam5007 emission-line luminosity of $9.6\times
10^{40}~{\rm erg~s^{-1}}$ is comparable to the [\ion{O}{3}]
luminosities of WLRGs, although its integrated 0.1-100~GHz radio
luminosity is an order of magnitude lower than the least luminous
WLRGs in the \citet{t98} sample. 

The weakness of the broad, disk-like Fe~K$\alpha$ line compared to
Seyferts appears to be a hallmark of BLRGs as a class
\citep{w98,sem99,esm00}.  This difference is illustrated in Figure~6
of \citet{hse02}, where the locations of Seyfert galaxies and BLRGs in
the Fe~K$\alpha~EW - L_{\rm 2-10~keV}$ plane are compared. In the vast
majority of cases, BLRGs fall {\it below} the region occupied by
Seyferts, with Arp~102B representing the lowest-luminosity BLRG with
available data.  The weakness of the Fe~K$\alpha$ line of Arp~102B can
be understood if the inner accretion disk has the form of an ion torus
or ADAF, rather than a thin disk surrounded by a hot corona, as is
thought to be the case in Seyfert galaxies \citep*{hm93,hmg94}. The
geometry of an optically thin, vertically extended inner disk
illuminating a geometrically thin, dense outer disk reduces the
available solid angle of the fluorescing medium (the thin disk) by a
factor of 2 relative to a thin disk sandwiched by a hot corona
\citep*{ch89,z99} and results in a reduction of the the Fe~K$\alpha$
$EW$ by the same factor. This picture is also appealing because it
explains some of the other properties of Arp~102B \citep[e.g., the
optical double-peaked emission lines and the shape of the SED;
see][]{ch89} and because it may be applicable to other low-luminosity
BLRGs \citep{esm00}. For example, 3C~332 is a dead-ringer for
Arp~102B: not only does it have a highly-absorbed X-ray spectrum and
an X-ray luminosity similar to that of Arp~102B \citep{cf95}, but it
also sports optical double-peaked emission lines and metastable
\ion{Fe}{2} absorption lines in the UV.

The above discussion refers to the extremely broad, disk-like
Fe~K$\alpha$ lines found in the \asca\ spectra of Seyfert galaxies.
However, recent observations of bright Seyfert galaxies at high
spectral resolution (with the \chandra\ grating spectrometers) or at
high $S/N$ (with the \xmm\ CCD cameras) have shown that the lines can
be decomposed into a broad, disk-like component and a narrower,
bell-shaped component. Examples include NGC~3783 \citep{k02}, NGC~5548
\citep{yaqoob02a}, and NGC~3516 \citep{tur02,netzer02}. The narrower
component has a width of a few thousand \kms\ and can be plausibly
attributed to the ``broad-line region,'' which is the source of the
broad, optical emission lines. The $EW$ of the narrower component is
typically around 90--130 eV. Such lines would not be detectable in the
\asca\ spectrum of Arp~102B, since their $EW$s are just below the
detection limit.

\subsection{Constraints on the Absorber, Implications, and Future Prospects}

In summary, we find that simple, single-zone models for the absorber
require high densities of $n\geq 10^{11}~{\rm cm^{-3}}$. However, not
all UV absorption lines can be produced at the same ionization
parameter: the Fe and Mg lines require an ionization parameter in
the range $10^{-3.5} < U < 10^{-2.5}$, while the Si, Al, C, and H
lines require a higher ionization parameter of $10^{-3.0} < U <
10^{-1.5}$ (the two regions of parameter space do not overlap,
however, as illustrated in Figure~\ref{fig_grid}).  The models that
reproduce the $EW$s of the Si, Al, C, and H lines predict Mg and Fe
line $EW$s that are much lower than the observed values, and {\it
vice-versa}: the models that reproduce the $EW$s of the Mg and Fe
lines predict Si, Al, C, and H line $EW$s that are much lower than the
observed values. The models that explain the Mg-Fe absorber can also
explain the X-ray absorption observed in the {\it ASCA} and {\it
ROSAT} spectra. Including a ``far-IR bump bump'' in the SED has a
negligible effect on these conclusions.

The simple, single-zone models that we have considered, although
partially successful, do not provide a fully satisfactory description
of the absorber. One possible way out of this difficulty would be to
adjust the abundances of Mg and Fe relative to the other elements by a
factor of a few.  However, more sophisticated models are probably
necessary and higher-resolution spectra are sorely needed to constrain
them and guide their development.  One possible effect that an
improved model could incorporate is radiative transfer in an
accelerating medium, since it may play a role in determining the
strengths of the absorption lines. This could be an important effect,
if the absorber has the form of an outflowing wind.  Another
possibility relevant to the outflowing wind scenario is that the
absorber spans a range of densities or that it contains pockets or
layers of higher or lower density gas.  Such a structure could lead to
the two phases that are apparently needed to explain the strengths of
the observed absorption lines. Different phases or layers of the
absorber may reveal themselves in the form of ``components'' of the
line profiles in high-resolution UV spectra. In this sense, our
preferred model for the Mg-Fe absorber is reminiscent of the model of
\citet{kraemer01} for NGC~4151: their high-resolution spectra allowed
them decompose the absorber into distinct kinematic components, one of
which was responsible for the metastable \ion{Fe}{2} absorption
lines. Their photoionization model for this particular component
required a high density of $n\geq 10^{9.5}~{\rm cm}^{-3}$, an
ionization parameter of $U=10^{-1.8}$, and a high column density of
$N_{\rm H}\approx 3\times 10^{21}~{\rm cm}^{-2}$. Their interpretation
was that this component of the absorber was the closest to the
ionizing source.

Examining the model results from a broader perspective and comparing
them with results of detailed models of absorption lines in Seyfert
galaxies, we find that a range of ionization parameters in the
absorbing gas is not at all uncommon; in fact, it is the rule rather
than the exception. This conclusion follows from studies of the UV and
X-ray absorption lines in bright Seyfert galaxies such as NGC~3783
\citep{k01,blustin02}, NGC~5548 \citep{crenshaw99,kaastra02}, NGC~4051
\citep{col01}, IRAS~13349+2438 \citep{sako01}, NGC~3516
\citep{kraemer02a}, and Mrk~509 \citep{yaqoob02b,kraemer02b} using
high-resolution spectra. More specifically, the UV absorption lines,
and sometimes even the X-ray lines, can be decomposed into several
kinematic components, which require gas with a wide range of
ionization parameters (this is sometimes obvious from a simple
comparison of the line profiles, without the need for detailed
modeling). In other cases a wide range of ionization is infered with
the help of photoionization models: such models cannot reporduce the
strengths of all of the observed lines under the assumption of a
uniform absorber with a single ionization parameter.  What is
different between Arp~102B and Seyfert galaxies, however, is the
presence of metastable \ion{Fe}{2} absorption lines and the
requirement for high densities to explain them. As we have noted in
previous sections, such lines are rare in Seyfert galaxies.

Taking the properties of the absorber in Arp~102B inferred from the
models at face value, we may deduce its distance from the source of
ionizing radiation. We find that the absorber distance is related to
the model parameters via $r=7\times
10^{16}\,(U_{-3}\,n_{11})^{-1/2}$~cm (where $U_{-3}=U/10^{-3}$ and
$n_{11}=n/10^{11}\;{\rm cm}^{-3}$).  Since $U = 10^{-3.0}$ and $n =
10^{11}~{\rm cm}^{-3}$ represent the lowest allowed photon flux
incident on the face of the absorber, the absorber distance should be
less than $7\times 10^{16}$~cm. We can recast this limit in terms of
the gravitational radius, $r_{\rm g} \equiv GM_{\bullet}/c^2$ (where
$M_{\bullet}$ is the mass of the central black hole), as $r/r_{\rm g}
\lsim 5,000\; (U_{-3}\,n_{11})^{-1/2}\; M_8^{-1}$ (where $M_8 =
M_{\bullet}/10^8\; M_{\odot}$). This distance is slightly larger than
the size line-emitting portion of the disk, inferred from fits to the
profiles of the broad Balmer emission lines, which extends over the
range $r/r_{\rm g}=500 - 1,000$ \citep{chf89,ch89}. We may also
estimate the thickness of the absorber along the line of sight as
$\delta r \sim N_{\rm H}/n$, which gives $\delta r \lsim 3\times
10^{10}$~cm, or $\delta r/r_{\rm g} \lsim 2\times 10^{-3}\;M_8^{-1}$,
assuming that $n\ge10^{11}\;{\rm cm}^{-3}$. Comparing the \ion{Fe}{2}
absorber in Arp~102B with those in luminous quasars, we find it to be
quite different: it is more dense and located closer to the ionizing
source. In comparison, the \ion{Fe}{2} absorber in Q~0059--2735 has
$10^6~{\rm cm^{-3}} < n < 10^8~{\rm cm^{-3}}$ and $2.4\times 10^5\;
M_8^{-1} \lsim r/r_{\rm g} \lsim 1.3\times 10^6\; M_8^{-1}$
\citep{wcp95}, while in the FIRST quasars studied by
\citet{dek01,dek02} the \ion{Fe}{2} absorber is characterized by
$n\sim 10^3-10^5~{\rm cm^{-3}}$ and $r/r_{\rm g}\sim 10^8\; M_8^{-1}$.

Independently of the model results, there are two more hints afforded
by the data: (1) the absorbing material is either outside of the
broad-line region or mixed in with it, since the \ion{Mg}{2}
absorption lines are deeper than the broad \ion{Mg}{2} {\it emission}
lines, and (2) the dust-to-gas ratio in the absorber is rather low,
since for a Galactic dust-to-gas ratio we would expect a reddening of
$E(B-V)=0.43$, which would strongly attenuate the broad UV emission
lines, contrary to what is observed \citep{h96}.  Putting all the
clues together, it is plausible to think of the absorber as thin,
dense sheets or filaments embedded in an outflowing wind overlaying
the outer accretion disk, which is thought to be the source of the
broad, double-peaked emission lines of Arp~102B. This outflowing wind
may also be the source of the broad UV high-ionization and Ly$\alpha$
emission lines \citep[see the general discussion by][and references
therein, and the specific discussion of Arp~102B by Halpern et
al. 1996]{cd89}. A very similar scenario may apply to other AGNs of
relatively low luminosity, which have many properties in common with
Arp~102B, including \ion{Fe}{2} absorption lines in their UV spectra,
for example the BLRG 3C~332 and the LINER NGC~1097 \citep{h97,e02}.

\acknowledgements

We are grateful to Gary Ferland for his patience and invaluable help
with the photoionization code Cloudy.  We also thank Chris Churchill for
his help in developing some of the software tools for this work and
the anonymous referee for thoughtful comments and suggestions. This 
work was supported by NASA through grants NAG5-10817 and NAG5-8369.

\clearpage

\def\aj{\rm{AJ}}                   
\def\araa{\rm{ARA\&A}}             
\def\apj{\rm {ApJ}}                
\def\apjl{\rm{ApJ}}                
\def\apjs{\rm{ApJS}}               
\def\apss{\rm{Ap\&SS}}             
\def\aap{\rm{A\&A}}                
\def\aapr{\rm{A\&A~Rev.}}          
\def\aaps{\rm{A\&AS}}              
\def\mnras{\rm{MNRAS}}             
\def\nat{\rm{Nature}}              
\def\pasj{\rm{PASJ}}    	   
\def\procspie{\rm{Proc.~SPIE}}     

\clearpage


\clearpage
\begin{deluxetable}{llclcl}
\tablenum{1}
\tablewidth{5.2in}
\tablecolumns{6}
\tablecaption{Exposure Times and Count Rates}
\tablehead{
\colhead{} &
\colhead{Observation} &
\colhead{Exposure} &
\colhead{Source} &
\colhead{Background} &
\colhead{Energy} \\
\colhead{} &
\colhead{Date} &
\colhead{Time} &
\colhead{Count Rate} &
\colhead{Count Rate} &
\colhead{Range} \\
\colhead{Instrument} &
\colhead{(UT)} &
\colhead{(ks)} &
\colhead{(s$^{-1}$)} &
\colhead{(s$^{-1}$)} &
\colhead{(keV)} 
}
\startdata
{\it ROSAT} PSPC  & 1991 Oct. 24 & \phantom{0}9.98 & 0.33$\pm$0.02   & 0.091 & 0.2--2.5  \\
{\it ASCA} SIS\,0 & 1999 Feb. 19 & 50.75 & 0.268$\pm$0.003 & 0.041 & 0.6--8.0  \\
{\it ASCA} SIS\,1 & 1999 Feb. 19 & 50.66 & 0.171$\pm$0.002 & 0.025 & 1.2--8.0  \\
{\it ASCA} GIS\,2 & 1999 Feb. 19 & 55.08 & 0.165$\pm$0.002 & 0.014 & 0.9--10.0 \\
{\it ASCA} GIS\,3 & 1999 Feb. 19 & 55.06 & 0.207$\pm$0.002 & 0.014 & 0.9--10.0 \\
\enddata
\end{deluxetable}

\begin{deluxetable}{lcccc}
\tablenum{2}
\tablewidth{6.0in}
\tablecolumns{5}
\tablecaption{Parameters of Best-Fitting Continuum Models \tablenotemark{a}}
\tablehead{
& \multicolumn{3}{c}{\it ASCA} & \colhead{\it ROSAT} \\
\noalign{\vskip -8 pt}
& \multicolumn{3}{c}{\hrulefill} & \colhead{\hrulefill} \\
\colhead{} &
\colhead{Simple} &
\colhead{Broken} &
\colhead{Power Law Plus} &
\colhead{Simple} \\
\colhead{Parameter} &
\colhead{Power Law} &
\colhead{Power Law} &
\colhead{Reflection \tablenotemark{b}} &
\colhead{Power Law} 
}
\startdata
Power Law Photon Index, $\Gamma$                        & $1.58\pm0.04$ & $1.6^{+1.0}_{-0.6}$ \tablenotemark{c} & $1.59^{+0.10}_{-0.07}$ & $< 2.2$ \\
Absorbing Column, $N_{\rm H}~(10^{21}~{\rm cm}^{-2})$   & $2.8\pm0.3$   & $2.7^{+1.4}_{-1.0}$ & $2.9^{+0.5}_{-0.4}$ & $1.0^{+4.0}_{-0.8}$ \\
Flux Density at 1~keV ($\mu$Jy) \tablenotemark{d}       & $1.7\pm0.2$   & $1.6\pm0.1$         & $1.7\pm0.2$       & $0.8^{+0.2}_{-0.1}$   \\
Break Energy (keV)                                      & \dots         & 2.4 \tablenotemark{e} & \dots           & \dots \\
High-Energy Photon Index, $\Gamma_2$                    & \dots         & $1.59^{+0.03}_{-0.08}$ & \dots          & \dots \\
Reprocessor Solid Angle, $\Omega/2\pi$                  & \dots         & \dots               & 0.4               & \dots \\
Reprocessor Inclination Angle, $i~(^{\circ})$           & \dots         & \dots               & 10                & \dots \\
Power Law Cutoff Energy (keV)                           & \dots         & \dots               & 100 (fixed)       & \dots \\
Reduced $\chi^2$ and d.o.f.                             & 1.053/279     & 1.060/277           & 1.058/277         &       \\
\sidehead{}
Flux ($10^{-11}~{\rm erg~s^{-1}~cm^{-2}}$) \tablenotemark{d,f} & $1.2\pm0.2$ & \dots            & \dots             & $0.13^{+0.05}_{-0.09}$ \\
Luminosity ($10^{43}~{\rm erg~s^{-1}}$) \tablenotemark{d,f}    & $3.1\pm0.6$ & \dots            & 100               & $0.3^{+0.1}_{-0.2}$ \\
Bandpass (keV)                                               & 2--10       & \dots            & \dots             & 0.5--2.0 \\
\tablenotetext{a\;}{All error bars correspond to 90\% confidence limits}
\tablenotetext{b\;}{Abundances of heavy elements were held fixed to their Solar values}
\tablenotetext{c\;}{The low-energy power-law index, below the break energy}

\tablenotetext{d\;}{Derived from the normalization of the spectrum. If
$N_{\rm E}$ is measured in ${\rm photons~s^{-1}~cm^{-2}~keV^{-1}}$,
then $f_{\nu}(1\;{\rm keV})=663\;N_{\rm E}~\mu{\rm Jy}$. In the case
of {\it ASCA} the uncertainty is dominated by uncertainties in the
absolute sensitivities of individual instruments. In the case of {\it
ROSAT} the normalization was determined after assuming that
$\Gamma=1.6$, with the uncertainty depending only on the uncertainty
in the total number of counts}

\tablenotetext{e\;}{The break energy was constrained to lie between 1 and 4~kev based
on the results of Wo\'zniak et al. (1998).}

\tablenotetext{f\;}{Fluxes are not corrected for absorption, while
luminosities are.}

\enddata
\end{deluxetable}


\clearpage
\begin{deluxetable}{lcccl}
\tablenum{3}
\tablewidth{6.5in}
\tablecolumns{5}
\tablecaption{Measured Properties of UV Absorption Lines}
\tablehead{
\colhead{} &
\multicolumn{2}{c}{Rest $EW$ (\AA)} &
\colhead{} &
\colhead{} \\
\noalign{\vskip -8 pt}
\colhead{Ion and} &
\multicolumn{2}{c}{\hrulefill} &
\colhead{FWHMin\tablenotemark{b}} &
\colhead{} \\
\colhead{Transition} &
\colhead{Measured} &
\colhead{Corrected\tablenotemark{a}} &
\colhead{~~(km~s$^{-1}$)~~} &
\colhead{Notes\tablenotemark{c}}
}
\startdata
\lion{Si}{3}{1206}       & $1.1 \pm 0.2$ & $1.8 \pm 0.3$ & 430   & $f_{\rm C}\approx 0.6$ \\
\ion{H}{1}~Ly$\alpha\;\lambda$1216   & $1.9 \pm 0.4$ & $2.4 \pm 0.5$ & 500   & $f_{\rm C}\approx 0.8$ \\
\lion{Si}{2}{1260}       & $< 1.0$       & $< 1.7$       & 500   & assumed FWHMin; assumed $f_{\rm C}=0.6$  \\
\lion{C}{2}{1334}        & $0.9 \pm 0.2$ & $1.5\pm 0.2$  & 290   & assumed $f_{\rm C}=0.6$\\
\lion{Al}{2}{1671}       & $< 1.4$       & $<2.3$        & 500   & assumed FWHMin; assumed $f_{\rm C}=0.6$ \\
\lion{Fe}{2}{2599}       & 3.6           & \dots         & \dots & Large uncertainty due to severe blending \\
\lion{Fe}{2}{2585}       & 2.5           & \dots         & \dots & Large uncertainty due to severe blending \\
\llion{Fe}{2}{2562,2563} & 4.4           & \dots         & \dots & Combined $EW$ of doublet \\
\lion{Mg}{2}{2796}       & $3.9 \pm 0.6$ & $8 \pm 1$ & 580 & $f_{\rm C}\approx 0.5$  \\
\lion{Mg}{2}{2803}       & $4.0 \pm 0.7$ & $8 \pm 1$ & 580 & $f_{\rm C}\approx 0.5$  \\
\lion{Mg}{1}{2853}       & $1.8 \pm 0.1$ & $4.5 \pm 0.3$ & 400 & $f_{\rm C}\approx 0.4$ \\
\tablenotetext{a}{$EW$ corrected for the effect of partial coverage, wherever 
                  applicable, as discussed in the text.}

\tablenotetext{b}{The uncertainty in the FWHMin is typically
50~km~s$^{-1}$}

\tablenotetext{c}{$f_{\rm C}$ is the coverage fraction of the
background source(s), as discussed in \S6.1 of the text.  For
consistency with other lines, $f_{\rm C}=0.6$ is assumed in cases when
it cannot be estimated. ${\rm FWHMin}=500\kms$ was assumed in the
determination of $EW$ upper limits.}

\enddata
\end{deluxetable}

\clearpage
\appendix

\section{The X-Ray Spectrum of MS~1718.6+4902}

\begin{figure*}[h]
\begin{minipage}[t]{3.5in}
\centerline{\psfig{width=4.3in,figure=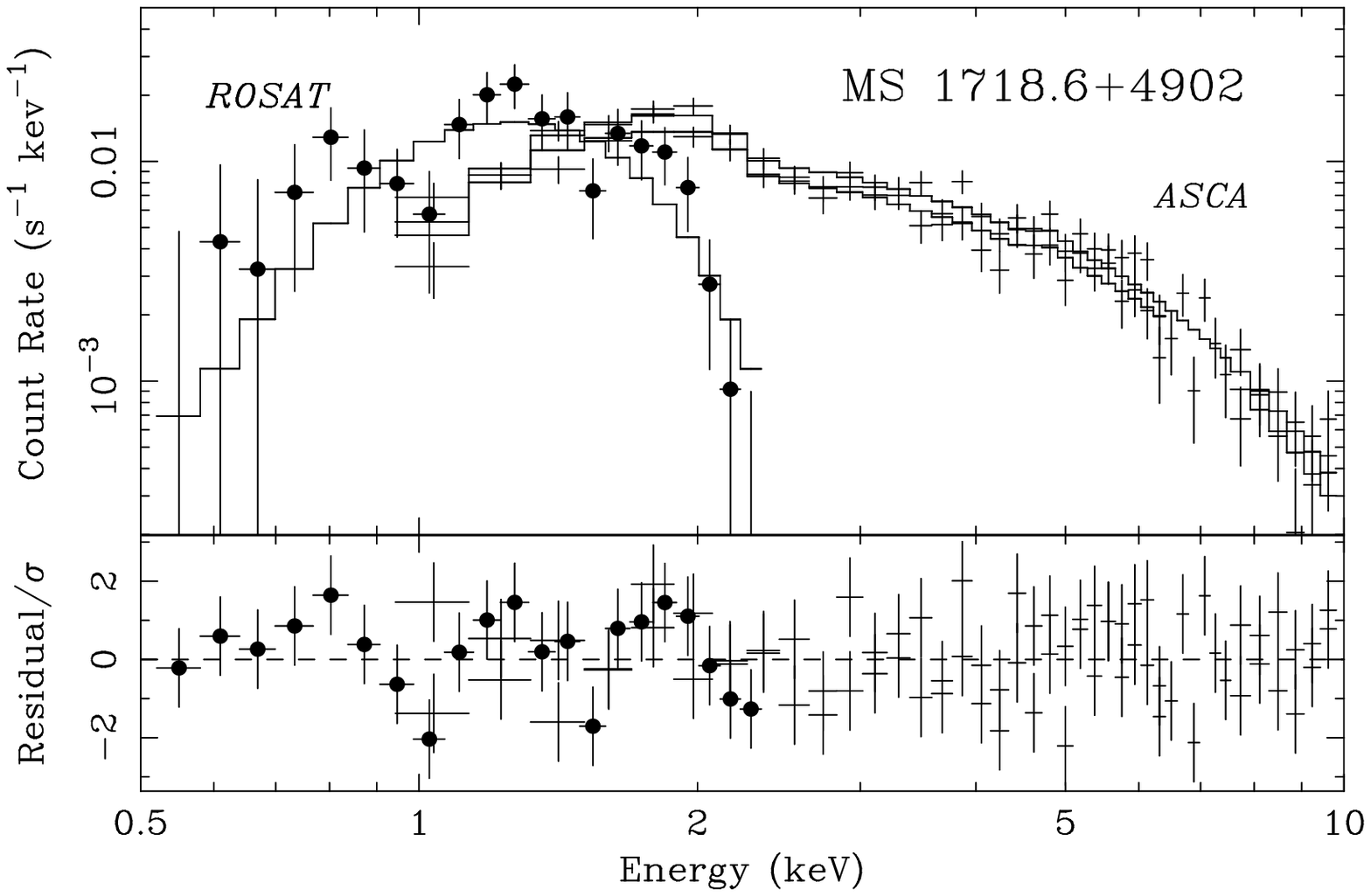}}
\figcaption{{\it Upper Panel:} Spectra of MS~1718.6+4902 from the {\it
ASCA} GIS and {\it ROSAT} PSPC with the best-fitting model
superposed. The model is an appropriately redshifted, heavily absorbed
power law with parameters as given in the text. {\it Lower Panel:}
Residuals from the subtraction of the best-fitting model from the
data, scaled by the error bar at each point.\label{fig_app_spec}}
\end{minipage}
\hskip 0.3truein
\begin{minipage}[t]{3.5in}
\centerline{\psfig{width=4.3in,figure=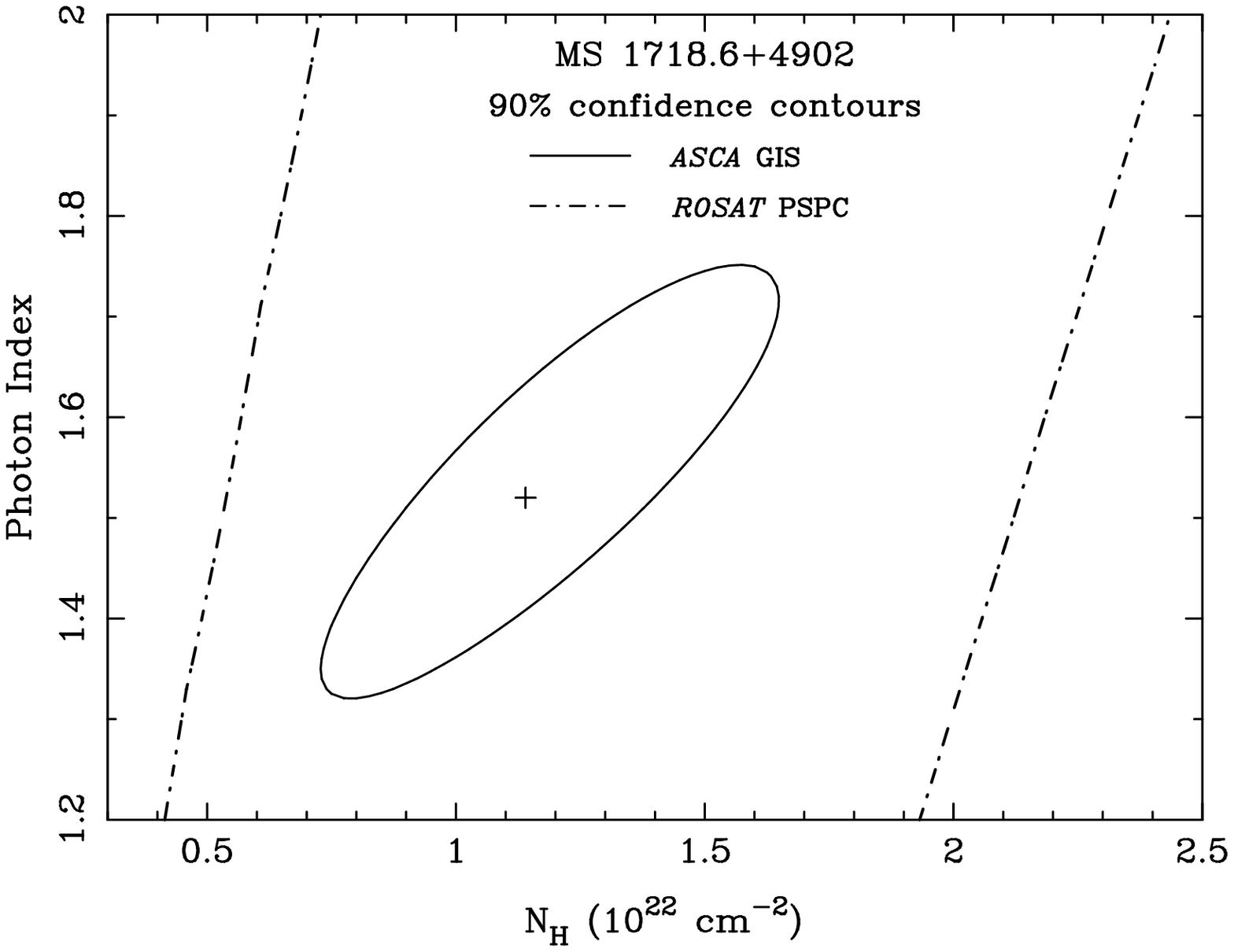}}
\figcaption{The 90\% confidence contours from the power-law model fit to the
spectra of MS~1718.6+4902. The solid lines shows the constraints derived from
the {\it ASCA} GIS spectra, while the dot-dashed line shows the constraints
derived from the {\it ROSAT} PSPC spectrum.\label{fig_app_cont}}
\end{minipage}
\end{figure*}

The galaxy MS~1718.6+4902 was detected as an X-ray source in the {\it
Einstein} Medium Sensitivity Survey (EMSS; Maccacaro et
al. 1991\nocite{m91}) and identified as an AGN at $z=0.198$ by
\citet{s91}. It is close enough to Arp~102B that it fell in the field
of view of both the {\it ROSAT} PSPC and the {\it ASCA} GIS and it was
detected by all of these instruments with net count rates of
0.02~s$^{-1}$ and 0.04~s$^{-1}$ respectively. We extracted an analyzed
the spectra of this object as described in \S2 and \S4. We find that
the {\it ASCA} GIS spectra are well described by a heavily absorbed
power law (appropriately redshifted) of photon index
$1.5^{+0.2}_{-0.1}$ and column density of $(1.2\pm0.3)\times
10^{22}~{\rm cm}^{-2}$ (the reduced $\chi^2$ is 1.085 for 72 degrees
of freedom). The observed 2--10~keV flux implied by this model is
$2.8\times 10^{-12}~{\rm erg~s^{-1}~cm^{-2}}$ and the intrinsic
luminosity, after correcting for absorption, is $5\times 10^{44}~{\rm
erg~s^{-1}}$. This absorbing column must be intrinsic to the AGN since
the Galactic column, inferred from the reddening reported by
\citet{sfd98}, is only $1.4\times 10^{20}~{\rm cm^{-2}}$. The {\it
ROSAT} spectrum, whose $S/N$ is relatively poor, is also consistent
with this model, although with a normalization that is a factor of
40\% higher than that of the {\it ASCA} spectra, presumably due to
variability of the source. In Figure~\ref{fig_app_spec} we show the
spectra with the best-fitting model superposed, while in
Figure~\ref{fig_app_cont} we show the 90\% confidence contours in the
photon index--column density plane.

\end{document}